\documentclass[reprint,pra,twocolumn,showpacs,superscriptaddress,aps,longbibliography]{revtex4-2}

\usepackage{graphicx}
\usepackage{epstopdf}
\usepackage{algorithm}
\usepackage{algpseudocode}
\usepackage[utf8]{inputenc}
\usepackage[T1]{fontenc}

\usepackage{lmodern}
\usepackage[english]{babel}

\usepackage{ae}
\usepackage{siunitx}

\usepackage{amsmath,amssymb,natbib,bm}
\usepackage{psfrag}
\usepackage{subfigure}
\usepackage{amsthm}
\usepackage{bbm}
\usepackage{algorithm}
\usepackage{algpseudocode}
\usepackage{listings}
\usepackage{multirow}
\usepackage{braket}
\usepackage{booktabs}
\usepackage{graphicx}
\usepackage[colorlinks]{hyperref}
\hypersetup{%
        plainpages=true,
        breaklinks=true,
        hypertexnames=false,
        pageanchor=true,
        colorlinks=true,
        linkcolor={blue},
        citecolor={blue},
        urlcolor={blue},
        anchorcolor={black}
      }

\usepackage{mleftright} 

\newcommand{\figref}[1]{\mbox{Fig.~\ref{#1}}}
\newcommand{\tabref}[1]{\mbox{Table~\ref{#1}}}
\newcommand{\secref}[1]{\mbox{Sec.~\ref{#1}}}

\newcommand{\appref}[1]{\mbox{Appendix~\ref{#1}}}
\renewcommand{\eqref}[1]{\mbox{Eq.~(\ref{#1})}}

\newcommand{\figpanel}[2]{Fig.~\hyperref[#1]{\ref*{#1}(#2)}}
\newcommand{\figpanels}[3]{Fig.~\hyperref[#1]{\ref*{#1}(#2)-(#3)}}
\newcommand{\figpanelNoPrefix}[2]{\hyperref[#1]{\ref*{#1}(#2)}}

\begin{document}

\title{Comparing and learning figures of merit for quantum circuit compilation}

\author{Harshdeep Singh}
\email{harshdeep.singh@chalmers.se}
\affiliation{Department of Microtechnology and Nanoscience, Chalmers University of Technology, 41296 Gothenburg, Sweden}

\author{Marvin Richter}
\affiliation{Department of Microtechnology and Nanoscience, Chalmers University of Technology, 41296 Gothenburg, Sweden}

\author{Mats Granath}
\affiliation{Department of Physics, University of Gothenburg, 41296 Gothenburg, Sweden}

\author{Anton Frisk Kockum}
\email{anton.frisk.kockum@chalmers.se}
\affiliation{Department of Microtechnology and Nanoscience, Chalmers University of Technology, 41296 Gothenburg, Sweden}


\begin{abstract}

To make quantum algorithms executable on a particular quantum device, they need to be compiled into circuits (sequences of quantum gates) that respect constraints of the quantum hardware. This compilation usually involves multiple steps, where many hardware-compatible circuits are generated, and the best circuit is selected. To say which circuit is best, the quality of a circuit (how well it executes the quantum algorithm) is generally quantified by a \textit{figure of merit} (FoM). For FoMs, there is a trade-off between ease of calculation and accuracy in predicted execution quality. Commonly used FoMs, e.g., the number of gates, circuit depth, etc., are easy to evaluate, but do not directly capture the effects of circuit structure and noise. On the other end of the spectrum are FoMs that require full circuit execution and take a prohibitively long time to evaluate. One example is the probability of successful trials (PST), i.e., the probability of obtaining the initial state after running the quantum circuit followed by its inverse. Here, we investigate advantages and disadvantages of different FoMs, and formulate the properties of an ideal FoM. Based on our results, we propose wPST, a weighted version of the PST that accounts for individual qubits, not just the whole state. To quickly predict PST and wPST, we design machine learning models that take into account both the quantum circuit and quantum hardware data. In numerical simulations and experiments on quantum processors, we find that our machine learning-predicted FoMs outperform commonly used FoMs, increasing the correlation with the true PST or wPST by over \SI{50}{\percent}. To make our model useful for quantum compilers, we devise a two-step process to predict the wPST for non-transpiled quantum circuits: first, we predict the additional quantum gates required for the given quantum circuit, and then we predict the wPST, accounting for coherence times in the quantum device. Our findings enable improvements in quantum circuit compilation, and thereby in the execution quality for quantum algorithms.

\end{abstract}

\date{\today}

\maketitle


\section{Introduction}

Quantum computers harness principles of quantum mechanics, such as superposition and entanglement, enabling them to represent and manipulate exponentially large state spaces~\cite{Nielsen2000}. In this way, quantum computers can solve certain classes of problems more efficiently than classical computers~\cite{Dalzell2025}; potential applications include quantum chemistry~\cite{qintro3, cao_2019_quantum, McArdle2020}, quantum dynamics~\cite{pf2, Georgescu2014}, materials research~\cite{Bauer2020}, and drug discovery~\cite{drug1, drug2, drug3}.

While quantum hardware and software have developed rapidly over the past few years~\cite{Madsen2022, Bluvstein2024, Acharya2025, Ransford2026, Yin2026}, quantum computing is still in the so-called noisy intermediate-scale quantum (NISQ) era~\cite{chalchap1}. In this era, a defining architectural feature of many quantum processors is restricted qubit connectivity and varying noise levels across qubits. This means that multi-qubit interactions are only directly available between specific pairs of physical qubits (usually nearest neighbors on a two-dimensional grid of some sort), rather than arbitrarily between all qubits in the device, and that which qubits are used for an operation will influence the quality of the result. 


These constraints imposed by the quantum hardware result in significant challenges for the implementation of quantum algorithms. The performance of a quantum algorithm is not just determined by the algorithm itself, but also by how efficiently it is mapped onto the specific quantum processor. This mapping process is called quantum circuit compilation~\cite{compil1, compil2}. It translates a high-level quantum algorithm into a sequence of physical operations (gates) that can be executed on a specific quantum hardware architecture. Finding the optimal compilation is computationally (NP-)hard because the space of possible circuits grows combinatorially with the number of qubits and gates~\cite{Cowtan2019, compil3}. Additionally, hardware constraints and noise effects make exhaustive search even more impractical.

To address these challenges, it becomes essential to (i) develop heuristic methods for compilation and (ii) define quantitative metrics that can quickly (ideally, without running any experiments) and reliably assess the quality of a compiled quantum circuit. These latter metrics, known as \textit{figures of merit (FoMs)}~\cite{fomintro, fompaper}, should reflect the quality of the execution of the compiled quantum algorithm. Thus, FoMs differ somewhat from characterization and high-level benchmarks for quantum computers~\cite{Proctor2025, zimboras2025euquantumflagshipskey}, which generally are designed to say more about hardware performance under the assumption of optimal compilation.

Some commonly used FoMs are as simple as the number of gates in the circuit, or the circuit depth, and are thus quick to calculate directly from the circuit without running any experiments. Other FoMs can be exponentially difficult to evaluate, like the fidelity of the output state from the compiled circuit. The goal is to find the sweet spot: an FoM that can actually characterize the quantum circuit, and the effects of quantum hardware on execution, but also is computationally feasible. Since FoMs need to be calculated many times during compilation, preferred FoMs have so far skewed towards the simplest ones, but studies have shown that their correlation with algorithmic execution quality is lackluster~\cite{Kurniawan2024, fompaper, Dangwal2025}.

In this article, we compare and develop FoMs and methods to evaluate them, aiming to find a better FoM tradeoff to improve quantum circuit compilation. In doing so, we focus on the probability of successful trials (PST)~\cite{pstintro}, an FoM that in its exact version is prohibitively costly already for tens of qubits. It requires executing the quantum circuit followed by its inverse to measure the probability of returning to the initial state by this procedure. Here, we introduce a weighted PST (wPST), which accounts for the states of individual qubits and not just the whole state, to better measure circuit quality. We analyze and compare the performance of wPST against other commonly used FoMs across different quantum devices. 

To get around the expensive evaluation of PST and wPST, we train machine learning-based models to predict these FoMs. We test the machine learning models in both numerical simulations and in experiments on IBM quantum computers. In these tests, we find an improvement of over \SI{50}{\percent} in the correlation with actual PST and wPST, as compared to other available FoMs calculated directly from circuit properties. 

Then, to make our predictive model useful for practical quantum compilers, we devise a two-step framework capable of estimating the performance of non-compiled quantum circuits. In the first step, we estimate the additional quantum gates that a given circuit will require when mapped to the target hardware, taking into account the hardware's qubit connectivity. This step effectively predicts the overhead introduced by compilation, such as the number of additional SWAP gates or extra depth resulting from hardware constraints, without actually performing a full compilation. We can then update the feature vector in the machine-learning model with these predictions. The second step then uses this updated feature vector for the final wPST prediction, the result of which the quantum compiler will apply to progress in its circuit selection. We believe that such integration of our machine-learning-predicted high-precision FoMs into quantum-compilation pipelines will result in improved compilation results, and thereby higher execution quality for quantum algorithms across hardware platforms.

This article is organized as follows. In \secref{sec:fom}, we first introduce common FoMs. Then, we discuss their shortcomings and propose wPST as an improvement. In \secref{sec:methods}, we show the methodology of our work, including the construction of a feature vector from quantum circuits, the machine learning model we employ, and how we train and benchmark it. In \secref{sec:results}, we show and discuss results quantifying the performance of our machine-learning-predicted FoMs in both numerical simulations and experiments on IBM quantum hardware. We then, in \secref{sec:2step}, present a two-step process that makes the machine-learning-predicted wPST method useful in a quantum compiler. Finally, we conclude and give an outlook for future work in \secref{sec:conclusion}. We provide some additional details about methods, devices, and results in the appendices.


\section{Figures of merit: Quantifying the quality of quantum circuits}
\label{sec:fom}

Figures of merit (FoMs) in quantum compilation are quantitative measures used to evaluate the quality and performance of a compilation~\cite{fomintro, fompaper}. They offer a systematic way to compare different compiled versions of the same quantum algorithm across various compilation strategies, optimization techniques, and hardware-aware mappings. Within quantum circuit compilation, choosing an FoM holds particular significance because improvements in one aspect of a circuit can sometimes compromise another. For instance, minimizing the number of SWAP operations could result in a less favorable error distribution across qubits~\cite{pstintro, fompaper}. 

A good figure of merit should have the following qualities:
\begin{itemize}
    \item \textbf{Hardware-aware}. It should take into account information from both the quantum circuits and the quantum hardware. This includes information about the placement of gates, the gate errors, coherence times, and the coupling graph.
    \item \textbf{Scalable}. It should remain meaningful and computable with growing circuit size. Direct metrics like the fidelity between the results of noisy and ideal circuit executions are exponentially expensive to calculate~\cite{spintomo, 1181970}, and are therefore impractical to use as an FoM. 
    \item \textbf{Computationally efficient}. It should be fast to evaluate. This ensures that the FoM can be used within quantum compilers, which often require searching through a large number of circuits.
    \item \textbf{Correlated with fidelity}. It should have a strong correlation with the fidelity. This ensures that the final compiled circuit produces the correct results.
    \item \textbf{Discrimination power}. It should be able to distinguish between `good' and `bad' results clearly. An FoM should be able to quantify the amount of noise or errors in the system. Previous works~\cite{Aamlof2013, PhysRevA.89.012305} show that fidelity might not be the perfect FoM in this respect, failing to indicate where the errors occur; this is further highlighted at the end of \secref{sec:pst}.
\end{itemize}

In the following, we present the commonly used FoMs for circuit compilation and discuss their shortcomings, before introducing the wPST as a viable solution. We are broadly classifying these FoMs into two categories:
\begin{enumerate}
    \item \textbf{The usual suspects}. These are the most commonly used FoMs, which can be directly evaluated from the circuit, without requiring results from running the quantum circuit and performing measurements. While thus generally easy to evaluate, these FoMs provide limited information about the quality of circuit execution.
    \item \textbf{Measurement-based}. These are more specialized FoMs that require running quantum circuits and obtaining results from measurements. These FoMs can be distance-based, which require results from both the noisy device and an ideal simulator, or probability-based, which require results from the execution of an extended quantum circuit, but are self-sufficient as they do not need results from an ideal simulator.
\end{enumerate}
Among the measurement-based FoMs, we pay particular attention to PST, since it is the precursor of the wPST that we propose at the end of the section.


\subsection{The usual suspects}

Commonly used FoMs that can be found directly from looking at the quantum circuit include circuit depth, total gate count, and two-qubit gate count. In quantum compilation, plenty of work has been done with the aim of improving such FoMs, e.g., reducing the total number of gates or the depth of the compiled circuit~\cite{sabre, kharkov2022arlinebenchmarksautomatedbenchmarking, Salm2021}, in the hope that this will improve the quality of the results for the executed quantum algorithm. 


\subsubsection{Circuit depth}

Circuit depth is the length of the critical path in the circuit's directed acyclic graph (DAG), or equivalently, the minimum number of time steps measured in gates that is required to execute the circuit when gates acting on disjoint qubits are parallelized. This FoM can act as a proxy for the circuit execution time and is also closely linked to the circuit size and decoherence effects. 


\subsubsection{Gate count}

Gate count, especially the number of two-qubit gates, is often used as an indicator of noise accumulation in a quantum circuit, as multi-qubit gates are generally more error-prone than single-qubit gates. The number of two-qubit gates is also often used as a tracking device for the additional SWAPs used by the compiler to make sure the quantum circuit can be executed on the given coupling structure of the quantum hardware. 


\subsubsection{Hardware-aware alternatives}

The above quantum-circuit-based FoMs are straightforward to compute and widely utilized, but they fail to explicitly consider device-specific noise characteristics~\cite{mqt, fompaper}. To address this limitation, more hardware-aware FoMs have been proposed. These FoMs include metrics based on expected fidelity~\cite{mqt, compilfid}, as well as estimated success probability (ESP)~\cite{esp1, esp3}, which takes into account the qubit coherence times (relaxation times $T_1$ and dephasing times $T_2$). These metrics provide a more realistic assessment of circuit performance on a specific backend. Graph-based methods have also been explored for quantum circuit compilation~\cite{Bandic2023, Venturelli2019QuantumCC, mqt, graphcompil}.

Importantly, no single FoM fully captures the performance of a compiled quantum circuit. Different metrics emphasize various physical and computational aspects, and their relevance may vary depending on the algorithm, hardware, and experimental objectives. This diversity of FoMs raises fundamental questions about how to best evaluate and compare compiled circuits, especially in the NISQ regime, where trade-offs are inevitable.

\begin{figure*}
    \centering
    \includegraphics[width=\linewidth]{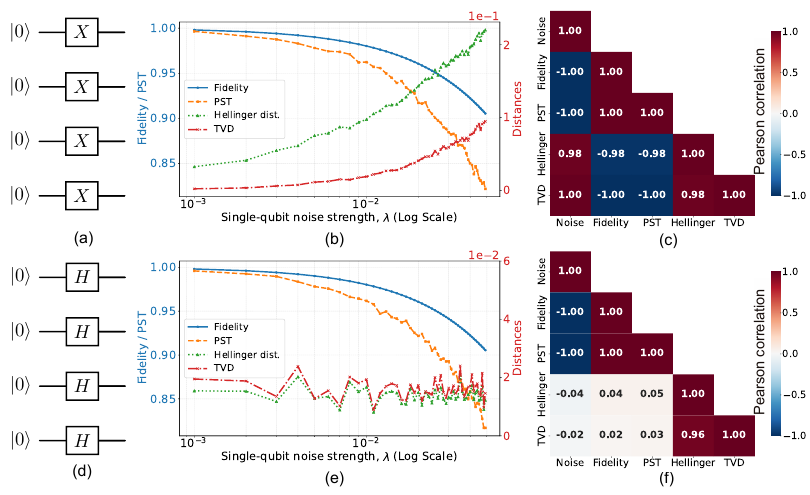}
    \caption{Measurement-based FoMs and their variation with noise for two four-qubit circuits.
    (a) A quantum circuit preparing the state $\ket{1111}$ in the absence of noise.
    (b) The values of fidelity, Hellinger distance, TVD, and PST for the circuit in (a), as a function of the depolarizing noise strength $\lambda$ [see \eqref{eq:DepolarizingNoiseDef}].
    (c) The Pearson correlations [see \eqref{eq:pcorrelation}] between the FoMs and the noise strength.
    (d) A quantum circuit preparing the equal superposition state $(\ket{0} + \ket{1})^{\otimes 4} = \frac{1}{4} \sum_{i=0000}^{1111} \ket{i}$ in the absence of noise.
    (e, f) Same as (b, c), but for the circuit in (d).}
    \label{fig:corr1}
\end{figure*}


\subsection{Measurement-based metrics}
To capture the complex interplay of gate errors, decoherence, and readout inaccuracies, we must look beyond the quantum circuit metrics and include the results from actually running the quantum circuit on the quantum device itself. In this subsection, we survey FoMs that are derived directly from the results of execution on quantum hardware, and propose wPST as an improved such FoM.


\subsubsection{Distance-based figures of merit}

When comparing quantum circuit outcomes, one way to quantify the performance is to rely on mathematical ``distances" to measure how far an observed output probability distribution diverges from the ideal output probability distribution~\cite{Mills2021, fompaper}.

\paragraph{Hellinger distance}

One such distance is the Hellinger distance, which is a statistical measure of dissimilarity between two probability distributions. This distance is widely used in statistics, information theory, and machine learning due to its well-defined metric properties and bounded nature. In the context of quantum computation, the Hellinger distance can provide an interpretable way to quantify the difference between an ideal and experimentally observed probability distribution of quantum circuits, making it a useful FoM for circuit compilation and execution quality. 

We here give a more precise definition of the Hellinger distance to facilitate the discussion. Let $P = \{p_i\}_{i=1}^n$ and $Q = \{q_i\}_{i=1}^n$ be two discrete probability distributions defined over the same finite sample space, such that
\begin{equation}
p_i \ge 0, \quad q_i \ge 0, \quad \sum_{i=1}^n p_i = \sum_{i=1}^n q_i = 1.
\end{equation}
The squared Hellinger distance between $P$ and $Q$ is defined as
\begin{equation}
H^2(P,Q) = \frac{1}{2} \sum_{i=1}^n \mleft( \sqrt{p_i} - \sqrt{q_i} \mright)^2
\end{equation}
and the Hellinger distance is given by
\begin{equation}
H(P,Q) = \frac{1}{\sqrt{2}} \mleft\| \sqrt{P} - \sqrt{Q} \mright\|_2,
\label{eq:hellinger}
\end{equation}
where $\sqrt{P} = (\sqrt{p_1}, \ldots, \sqrt{p_n})$ and $\|\cdot\|_2$ denotes the Euclidean norm.

Let $P^{\mathrm{ideal}}$ denote the ideal output probability distribution obtained from a noiseless simulation of a quantum circuit, and let $P^{\mathrm{comp}}$ denote the output distribution obtained from executing the compiled circuit on a target quantum device or from a noisy simulation. The Hellinger distance
$H\mleft(P^{\mathrm{ideal}}, P^{\mathrm{comp}}\mright)$
then provides a direct quantitative measure of how much the compiled circuit deviates from the ideal behavior.

The Hellinger distance can be used to rank compiled circuits according to their closeness to the ideal distribution. A shorter Hellinger distance indicates that the compilation process has preserved the logical structure of the original circuit better under hardware constraints and noise.


\paragraph{Total variation distance}

Similar to the Hellinger distance, one can also use the total variation distance (TVD) as an FoM. The TVD corresponds to the classical limit of the quantum trace distance. For two classical probability distributions $P$ and $Q$ defined over a finite sample space $\Omega$, the TVD is defined as
\begin{equation}
D_{\mathrm{TV}}(P,Q)
= \frac{1}{2} \sum_{x \in \Omega} \mleft| P(x) - Q(x) \mright|.
\end{equation}
This quantity measures the distinguishability between the two distributions and satisfies
\begin{equation}
0 \le D_{\mathrm{TV}}(P,Q) \le 1,
\end{equation}
with $D_{\mathrm{TV}}(P,Q) = 0$ if and only if $P = Q$, and $D_{\mathrm{TV}}(P,Q) = 1$ if and only if the distributions have disjoint support.  

While both the TVD and the Hellinger distance serve a similar operational purpose---providing a bounded $[0,1]$ metric to evaluate distribution similarity---they possess distinct mathematical properties. Fundamentally, the two metrics are linked by the following bounds:
\begin{equation}
H^2(P,Q) \le D_{\mathrm{TV}}(P,Q) \le \sqrt{2} H(P,Q).
\end{equation}
The primary difference lies in the sensitivity of these distances to probability magnitudes. The TVD evaluates the linear absolute difference between probabilities; it treats an error shift from $0.50$ to $0.51$ exactly the same as a shift from $0.00$ to $0.01$. In contrast, because the Hellinger distance evaluates the difference between the square roots of the probabilities ($\sqrt{p_i}$), it is non-linear and highly sensitive to changes in near-zero probabilities. 

In quantum computing, ideal circuits often produce sparse distributions where many basis states have a theoretical probability of zero. When executed on noisy hardware, decoherence and gate errors cause ``leakage'' into these states. The Hellinger distance penalizes these small, erroneous probabilities much more heavily than the TVD. Consequently, Hellinger distance is often preferred for hardware benchmarking, where capturing subtle noise signatures and penalizing low-probability anomalies is paramount~\cite{fompaper}.

\paragraph{Challenges for distance-based metrics}

While distance-based metrics are often used as an FoM for quantum circuit compilation, they face some fundamental challenges. For an $n$-qubit quantum circuit, both the ideal and noisy executions produce probability distributions over the computational basis $\{0,1\}^n$, which contains $2^n$ possible outcomes. Therefore, the computational complexity of calculating the Hellinger distance between two distributions $P$ and $Q$, as defined in \eqref{eq:hellinger}, scales as $\mathcal{O}(2^n)$, since the contribution from every computational basis state must be explicitly computed.

This exponential scaling constitutes a major bottleneck for data-driven quantum compilation. In supervised-learning frameworks, a large number of labeled training examples is typically required to learn robust compiler heuristics or performance predictors. If distance-based FoMs are used as training labels, the overall cost of data-set generation scales as
$ \mathcal{O} \mleft(m \cdot 2^n\mright)$,
where $m$ denotes the number of training circuits. Even for moderate values of $n$, this scaling renders data-set construction computationally infeasible.

Another major challenge is that distance-based metrics can be unreliable when the circuit produces a broad output distribution, rather than concentrating probability on a few states. As an illustration, consider two four-qubit circuits, one with a single output [\figpanel{fig:corr1}{a}, where the output is fixed as $\ket{1111}$] and one with multiple outputs [\figpanel{fig:corr1}{d}, where the final result is an equal superposition of all possible states, $(\ket{0} + \ket{1})^{\otimes 4}$]. 

The focus of the example in \figref{fig:corr1} is on how the distance-based metrics respond to increasing noise in each case. The noise strength of a depolarising channel $E$ implemented in Qiskit~\cite{depolarizing} can be controlled by varying the parameter $\lambda$:
\begin{equation}
    E(\rho) = (1-\lambda)\rho + \lambda \text{Tr} [ \rho ]\frac{I}{2^n} ,
\label{eq:DepolarizingNoiseDef}
\end{equation}
where $\rho$ is the density matrix, $n$ is the number of qubits, and $0\le \lambda \le 4^n/(4^n-1)$. Now, if we increase the single-qubit depolarising noise, the distance-based metrics have a very strong correlation to noise strength and fidelity when the final result of a circuit measurement is concentrated on one or very few states, as shown in \figpanels{fig:corr1}{b}{c}. However, when the measurement outcome is spread across many states, the distance-based metrics often show very low correlation to the noise strength and fidelity, as highlighted in \figpanels{fig:corr1}{e}{f}. It must be noted that for \figpanel{fig:corr1}{e}, with an infinite number of measurements, the distances would be zero, and the fluctuations are merely statistical. This is why we shift our focus to the FoM probability of successful trials (PST), which can be self-sufficient, scalable, and highly correlated to fidelity across a wide range of output distributions.


\subsubsection{Probability of successful trials}
\label{sec:pst}

The probability of successful trials (PST) is a FoM that quantifies how faithfully a quantum circuit is implemented on a given quantum device. It is based on the idea of \emph{circuit inversion}, or \emph{mirror circuits}, and provides a direct operational measure of how close the realized computation is to the intended unitary transformation; it has been shown to have a strong correlation with circuit fidelity~\cite{torchquantum}. Mirror circuits are often used as a benchmarking technique to evaluate the performance of quantum gates and to reveal and quantify the crosstalk errors in multi--qubit circuits~\cite{PhysRevLett.129.150502}.

Unlike metrics that compare full output distributions, PST focuses on a single well-defined reference outcome, making it particularly useful for benchmarking compilation quality in the presence of noise. Operationally, PST is the gate-based circuit analogue of the quantity introduced by Peres~\cite{peres1984} and later termed the \emph{Loschmidt echo} by Jalabert and Pastawski~\cite{jalabert2001}: both measure the overlap of a forward-evolved state with its time-reversed counterpart in the presence of imperfections.

We now define PST more precisely. Let $U$ be the unitary operator implemented by an ideal quantum circuit acting on $n$ qubits. Consider the initial state
\begin{equation}
\ket{\psi_0} = \ket{0}^{\otimes n}.
\end{equation}
In the absence of noise, applying $U$ followed by its inverse $U^{-1}$ yields
\begin{equation}
U^{-1} U \ket{\psi_0} = \ket{\psi_0}.
\end{equation}

In PST, we construct a \emph{verification circuit} by appending the inverse of the circuit to itself. In practice, the circuit $U$ is first compiled to a hardware-executable circuit $\tilde{U}$, and its inverse $\tilde{U}^{-1}$ is generated by reversing the gate sequence and inverting each gate. The full circuit is therefore $\tilde{U}^{-1} \tilde{U}$.

After executing this composite circuit on a quantum device and measuring in the computational basis, the PST is defined as the probability of obtaining the all-zero outcome:
\begin{equation}
\text{PST} = p(0^{\otimes n}),
\end{equation}
where $0^{\otimes n}$ denotes the bit string consisting of $n$ zeros.

In an experimental setting, the circuit is executed $N$ times (shots). Let $n_{0^{\otimes n}}$ denote the number of times the outcome $0^{\otimes n}$ is observed. The empirical PST is then
\begin{equation}
\text{PST} = \frac{n_{0^{\otimes n}}}{N}.
\end{equation}

However, PST also has limitations. The doubled circuit depth resulting from appending the inverse can exaggerate noise effects, and the metric does not distinguish which parts of the circuit contribute most to the error. Furthermore, PST does not tell us well how close the circuit was to performing perfectly; a single bit flip has the same effect as many bit flips on the PST value (they set it to zero). Indeed, this is a problem for fidelity as well, since the computational bases are orthogonal and a single bit flip immediately leads to a zero fidelity result; we note that fidelity has additional problems in assessing how close states are to each other~\cite{PhysRevA.89.012305}. For these reasons, PST is most informative when used alongside other FoMs, such as circuit depth, gate counts, and distribution-based distances.



\subsubsection{Weighted probability of successful trials}
\label{sec:wpst}

To overcome these drawbacks of PST, we introduce a variant of PST that takes into account how many zeros are measured (down to some limit), rather than just counting the number of all-zero measurement results. We call this FoM the \emph{weighted} probability of successful trials (wPST). 


\paragraph{Definition of wPST.}

Consider an experiment with many shots on the same verification circuit $\tilde{U}^{-1} \tilde{U}$ as in \secref{sec:pst} above, where the computational basis states for $n$ qubits, $\{\ket{0000..0}, \ket{00..01}, \ket{00..10}, ...\}$ were measured with probabilities $\{p_1, p_2, p_3, ...\}$, yielding classical bit strings $x_i \in \{ 0,1\}^n$. The weights, in the context of wPST, are defined as the number of non-flipped bits divided by the total number of bits,
%
\begin{equation}
    w_i = \frac{\#\,0\text{s in } x_i}{n} = 1 - \frac{h_i}{n} ,
    \label{eq:weights}
\end{equation}
where $h_i$ is the Hamming weight of $x_i$, i.e., the number of bits that were flipped away from the all-zero outcome.
For example, for the measurement result $0000$ in a four-qubit experiment, we assign weight $w=1$, while for $0001$ the weight would be $w=0.75$. Using these weights, one can essentially quantify the amount of `correct' result in the quantum measurement. 

The wPST can then be defined as
\begin{equation}
\mathrm{wPST} =
\sum_{i: w_i \ge w_{\rm th}} p_i w_i ,
\label{eq:wpst}
\end{equation}
so that wPST is a Hamming-weight-graded PST that assigns partial credit toward the overall correct states. To ensure that we only reward overall relatively correct results, we use a threshold weight ($w_{\rm th}$), which we set to 0.5 in the simulations and experiments later in the article. For states with $w<w_{\rm th}$, one is looking at a result where the final outcome has more incorrect bits than correct ones, and those cases should not be rewarded in the FoM calculation. 

The behavior of \eqref{eq:wpst} is most transparent in the case $w_{\rm th}=0$. Too see this, we write the weight of a bit string as an average of per-bit indicators,
\begin{equation}
    w_i = \frac{1}{n} \sum_{q=1}^n \mathbbm{1}\left[x_{i,q} = 0\right],
    \label{eq:wi}
\end{equation}
where $\mathbbm1[x_{i,q}=0]$ is an indicator function that counts the number of zeros in each measurement bit string, 
\begin{equation}
\mathbbm{1}[x_{i,q}=0]
=
\begin{cases}
1, & x_{i,q}=0,\\
0, & x_{i,q}=1.
\end{cases}    
\end{equation}
On exchanging the two sums in Eqs.~(\ref{eq:wpst}) and (\ref{eq:wi}), we obtain
\begin{equation}
\begin{split}
    \mathrm{wPST}_{w_{\rm th}=0}
    &= \sum_i p_i \frac{1}{n}\sum_{q=1}^{n}\mathbbm{1}\left[x_{i,q}=0\right] \\
    &= \frac{1}{n}\sum_{q=1}^{n} m_q \equiv \overline{m}_q .
\end{split}
\label{eq:wpstmarginals}
\end{equation}
where $m_q = \sum_i p_i\,\mathbbm{1}\left[x_{i,q}=0\right]$ is the single-qubit marginal probability that qubit $q$ is read as $0$. In other words, wPST is a \emph{linear} functional of the single-qubit marginals, whereas PST is a functional of the full joint distribution.  

The role of the threshold itself becomes clearer once the sum in \eqref{eq:wpst} is reorganized by Hamming weight. Let $k_i = n - h_i$ denote the number of zeros in $x_i$ and let 
\begin{equation}
    Q(k) = \sum_{i:\,k_i = k} p_i,
    \label{eq:shell}
\end{equation}
be the total probability of measuring any bit string with exactly $k$ zeros. Since $w_i$ depends on $x_i$ only through $k_i$, the threshold acts on the Hamming-weight shells, and with $k_{\rm th}=\lceil w_{\rm th} n\rceil$, where $\lceil. \rceil$ is the ceiling function that rounds up the product $w_{\rm th} n$ to the nearest integer, we obtain
\begin{equation}
\begin{split}
    \mathrm{wPST}_{w_{\rm th}}
    &= \sum_{i:\,w_i \ge w_{\rm th}} p_i w_i \\
    &= \sum_{k=k_{\rm th}}^{n}\ \sum_{i:\,k_i=k} p_i \frac{k}{n} = \frac{1}{n}\sum_{k=k_{\rm th}}^{n} k\, Q(k) .
\end{split}
\label{eq:wpsthamming}
\end{equation}
The threshold thus simply selects the lowest shell of bit strings that still receives partial credit, and wPST is the mean fraction of correct bits restricted to those shells. The two limiting cases interpolate between the FoMs discussed above: for $w_{\rm th}=0$ (i.e., $k_{\rm th}=0$) all shells contribute and we recover the marginals, $\mathrm{wPST}_0 = \overline{m}_q$, while for $w_{\rm th}=1$ (i.e., $k_{\rm th}=n$) only the all-zero string survives and $\mathrm{wPST}_1 = Q(n) = \mathrm{PST}$.


\paragraph{How wPST overcomes challenges for figures of merit.}

\begin{figure}
    \centering
    \includegraphics[width=\linewidth]{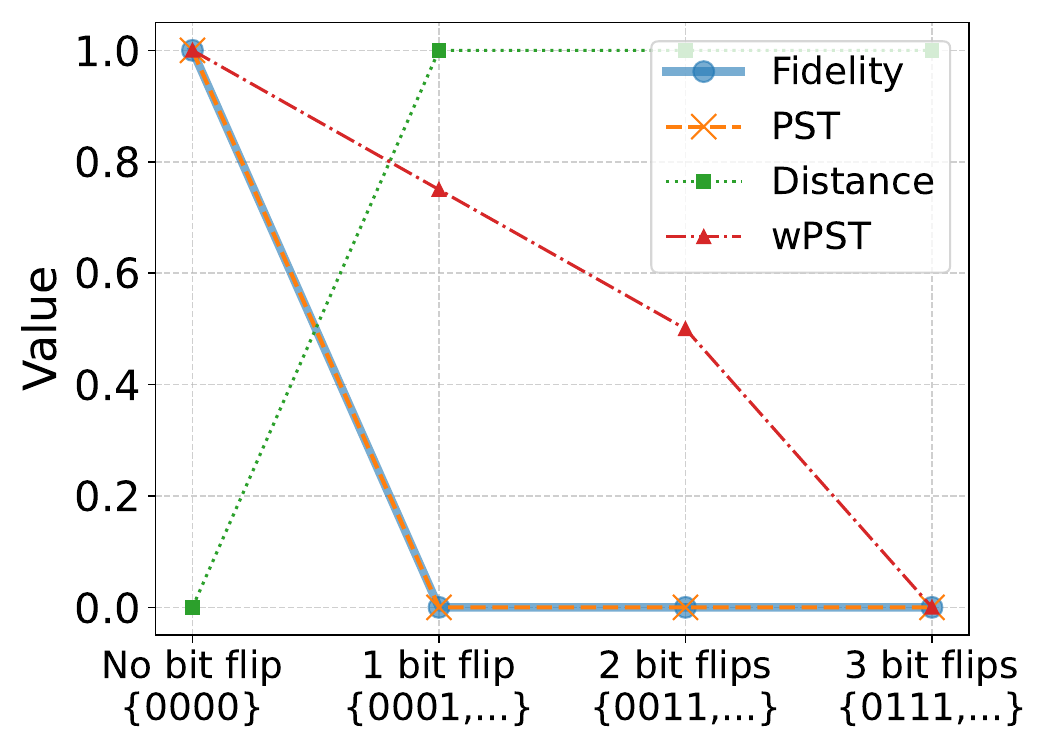}
    \caption{Values of fidelity, PST, distance-based metrics, and wPST for a four-qubit quantum circuit with increasing number of bit-flip errors.}
    \label{fig:fomerrors}
\end{figure}

To show how wPST resolves the major drawbacks of standard PST and distance-based FoMs, we present the results for these different FoMs in a simple example in \figref{fig:fomerrors}. Here, we consider a four-qubit quantum circuit under ideal and noisy conditions. 
If, in both cases, the measured outcome is the same (say, 0000), the fidelity, PST, and wPST are equal to 1, while the distance (both Hellinger and TVD) between the distributions is 0. Now, if there is a single bit flip, and the noisy result is of the form $\{ 0001, 0010, 0100, \ldots \}$, then fidelity and PST immediately drop to 0, while the distance increases to 1, but wPST is equal to 0.75, owing to the weights of the $0$s in the resulting string. In the case where there is more than one bit flip, the fidelity, PST, and distance remain unchanged, and thus cannot distinguish between the results of different noise levels. 
However, wPST provides a way to successfully distinguish between these different scenarios, as \figref{fig:fomerrors} shows; by \eqref{eq:wpsthamming}, each additional bit flip moves probability weight down one Hamming-weight shell and lowers the FoM by $1/n$.

In realistic scenarios, one needs to run a quantum circuit plenty of times to obtain the average distribution, and these single-shot errors can accumulate rapidly, rendering FoMs like PST useless. As explained in \eqref{eq:wpsthamming},  wPST is a \emph{linear} functional of the single-qubit marginals, whereas PST is a functional of the full joint distribution. This highlights the biggest difference between wPST and PST: the wPST is additive in the qubits rather than multiplicative. So, under an independent-error-per-qubit model with error rate $\epsilon$, the PST varies as
\begin{equation}
    \mathrm{PST}\approx(1-\epsilon)^n ,
    \label{eq:pstdecay}
\end{equation}
decaying exponentially in $n$ even for small $\epsilon$, while for wPST, we have 
\begin{equation}
    \mathrm{wPST}\approx (1-\epsilon) ,
    \label{eq:wpstdecay}
\end{equation}
decaying linearly with $\epsilon$ and independently of $n$. This behavior is explored further in Appendix~\ref{app:bitflip}, where we show that the number of measurements required to determine wPST scales better with the number of qubits than for PST.  

\begin{figure}
    \centering
    \includegraphics[width=\linewidth]{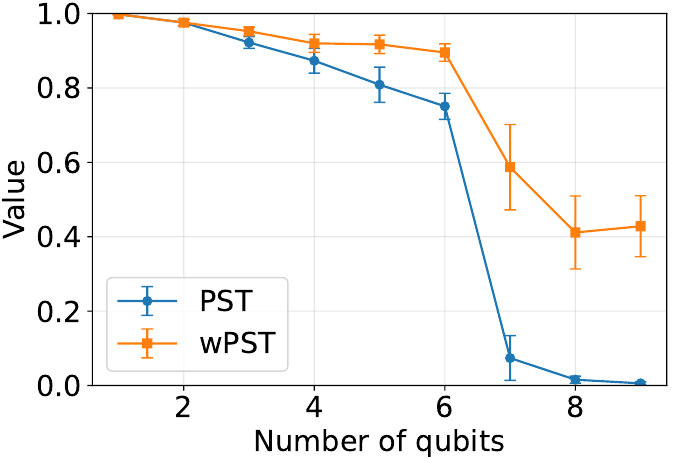}
    \caption{Mean values of PST and wPST for quantum circuits with increasing number of qubits with a simulator based on the IBM Miami device. For each qubit, 25 circuits were evaluated with a maximum depth of $5 \times n$ (pre-transpilation, $n=$~number of qubits). The error bars show $\pm 1$ standard deviation. }
    \label{fig:wpstpst}
\end{figure}

Therefore, for larger circuits, it would be very difficult to acquire the absolutely correct results ($0^{\otimes n}$) enough times to obtain a meaningful PST result. The majority of the time, one ends up with a few errors, and these results are not accounted for in PST calculations; the PST values drop rapidly with the system size, as shown in \figref{fig:wpstpst}. That figure shows the mean values of PST and wPST for quantum circuits with increasing number of qubits, using a simulator based on the IBM Miami device. 25 circuits for each qubit count and a maximum variable depth based on the number of qubits $n$ ($5 \times n$) were used. We observe that beyond six qubits, the PST values drop to near $0$, and have almost no deviation for the different circuits. At the same time, wPST retains the linear-in-$\epsilon$ behaviour inherited from \eqref{eq:wpstmarginals}, can still provide a range of values, and might actually provide some useful information in those cases.


\section{Methods}
\label{sec:methods}

While wPST has attractive properties when it comes to providing information about the quality of a quantum circuit, estimating the wPST (or PST) value is a computationally expensive task, especially for increasingly large quantum circuits. To accurately calculate wPST requires executing the quantum circuit on real hardware or a high-fidelity simulator, and performing a sufficiently large number of measurements to obtain statistically reliable estimates. This process can become prohibitively costly in the context of quantum-compilation workflows, where millions of potential circuits may need to be evaluated repeatedly during an optimization that ideally should take seconds. Consequently, direct computation of (w)PST for each candidate circuit is not scalable and poses a significant bottleneck.

To address this challenge, we develop a machine learning-based approach to approximate wPST quickly at low computational cost. Instead of relying on repeated circuit executions, we train a predictive model that learns the relationship between circuit characteristics, hardware properties, and the resulting wPST. Once trained, this model can provide fast and reasonably accurate estimates of this FoM, enabling efficient exploration of the compilation search space. This significantly reduces the computational overhead while retaining the ability to guide optimization decisions effectively.

\begin{figure*}
    \centering
    \includegraphics[width=\linewidth]{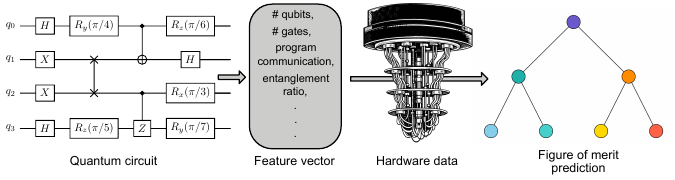}
    \caption{Workflow of our machine learning-based FoM prediction. The process involves constructing a feature vector from the quantum circuit and, after adding data from the quantum hardware, using an XGBoost model (based on decision trees) for training and prediction. 
    }
    \label{fig:pipeline}
\end{figure*}

In this section, we will show our pipeline for predicting both PST and wPST of quantum circuits by integrating circuit-level features with hardware-aware information. The overall framework, illustrated in Fig.~\ref{fig:pipeline}, begins with a quantum circuit as the input and progressively transforms it into a structured representation amenable to machine learning models. In the following, we show how to efficiently encode the quantum circuit and quantum hardware data into the input vector data, and then discuss the machine learning approach employed. 


\subsection{Encoding quantum circuits}
\label{sec:EncodingCircuits}

\begin{figure}
    \centering
    \includegraphics[width=\linewidth]{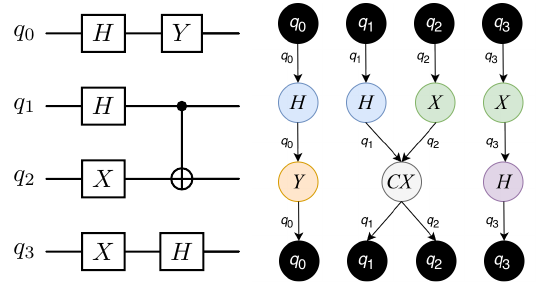}
    \caption{Directed acyclic graph (DAG, right) representation of a four-qubit quantum circuit (left).}
    \label{fig:dag}
\end{figure}

The first step in the process of estimating wPST for a quantum circuit involves extracting features that capture meaningful characteristics of the circuit. To make sure that feature extraction is efficient enough for a quantum compiler, the properties have to be carefully chosen. Given a quantum circuit $\mathcal{C}$ acting on $n$ qubits, we extract a set of structural, topological, and operational features that characterize its complexity and execution behavior after compilation, expanding on the features used in previous works~\cite{fompaper, mqt}. In extracting the features, we make use of the fact that the quantum circuit can be represented as a directed acyclic graph (DAG) $G = (V, E)$, as shown in \figref{fig:dag}, where
\begin{itemize}
\item Each node $v \in V$ corresponds to a quantum operation.
\item A directed edge $(u,v) \in E$ exists if operation $v$ depends on the output of operation $u$.
\end{itemize}

The features we derive from the circuit's DAG representation and from its gate-level description are listed below.
\begin{enumerate}
\item The total \emph{number of qubits}, $n$, used in the circuit.
\item The total \emph{number of quantum gates} in the circuit,
\begin{equation}
N_{\mathrm{g}} = |\mathcal{G}|,
\end{equation}
where $\mathcal{G}$ is the ordered set of all single- and multi-qubit gates in the circuit.
\item The \emph{circuit depth}, $D$, defined as the longest path of sequential layers that need to be executed in the full quantum circuit,
\begin{equation}
D = \max_{\pi \in \mathcal{P}} |\pi|,
\end{equation}
where $\mathcal{P}$ is the set of all gate paths when gates acting on disjoint qubits are parallelized.
\item The \emph{number of two-qubit (entangling) gates}, $N_2$. 
%
\item The total \emph{number of edges},
\begin{equation}
N_{\mathrm{edges}} = |E|.
\end{equation}
%
\item The length of the \emph{longest path} in the DAG,
\begin{equation}
L_{\max} = \max_{\pi \in \mathcal{P}} (|\pi| - 1),
\end{equation}
where $\pi$ is a directed path in $G$. 
%
\item The \emph{program communication} feature, $C_{\mathrm{prog}}$, defined as a normalized measure of qubit interactions within a quantum circuit. Let $\langle k \rangle$ denote the average degree of the DAG.  We normalize this quantity by the average degree of a complete graph with the same number of qubits, nodes, and edges, which is given by $\langle k_{\mathrm{complete}} \rangle = n - 1$. The program communication feature is therefore computed as
\begin{equation}
C_{\mathrm{prog}} = \frac{\langle k \rangle}{n - 1}.
\label{eq:ProgramCommunication}
\end{equation}
%
\item The \emph{entanglement ratio}, defined as the fraction of two-qubit gates in the circuit:
\begin{equation}
R_{\mathrm{ent}} = \frac{N_{2}}{N_{\mathrm{g}}}.
\end{equation}
%
\item The \emph{circuit parallelism}, which quantifies how efficiently gates are scheduled across available qubits:
\begin{equation}
P = \max\mleft(0, \frac{\frac{N_{\mathrm{g}}}{D} - 1}{n - 1}\mright).
\end{equation}
This normalized measure lies in $[0,1]$; higher values indicate greater simultaneous gate execution.
\item The \emph{critical depth}, $D_\mathrm{crit}$, of a quantum program, which quantifies the number of two-qubit interactions that lie along the \emph{critical path} of the circuit. The critical path is defined as the longest sequence of operations that must be executed sequentially due to qubit dependencies. Formally,
\begin{equation}
D_\mathrm{crit} = \frac{N_{e_{d}}}{N_2}
\end{equation}
where $N_{e_{d}}$ is the number of two-qubit gates along the longest path of the quantum circuit.  
%
\item The \textit{liveliness}, $L$, of a quantum circuit, which characterizes how well a program utilizes its qubits throughout its execution. During the execution of a quantum circuit, the qubits can either be \textit{active} (currently involved in gate operations) or \textit{idle} (waiting for the next operation). In ideal settings, the idle qubits can preserve their quantum state, but on the NISQ hardware, they can lose information due to amplitude damping, dephasing, etc. The liveliness is defined as 
\begin{equation}
    L = \frac{1}{nD} \sum_{i,j} A_{ij} ,
\end{equation}
where $A_{ij}$ is an activity indicator ($A_{ij}=1$ when qubit $i$ is active at the depth layer $j$).
\end{enumerate}

This feature vector is designed to capture a comprehensive set of structural and operational characteristics of a quantum circuit, enabling an informative and compact representation suitable for the machine learning model. Basic structural attributes such as the number of qubits, total gate count, and circuit depth provide a measure of the circuit’s scale and computational complexity. Graph-based features, like the number of edges and the longest path length, reflect the underlying interaction graph and the dependency structure between operations. 

Importantly, different subsets of features are tailored to capture distinct aspects of circuit behavior. For instance, the number of two-qubit gates and the entanglement ratio quantify the degree of multi-qubit interaction, which is closely tied to noise sensitivity and hardware constraints. Features such as parallelism and critical depth characterize the extent to which operations can be executed concurrently, thereby reflecting scheduling efficiency and temporal structure. Program communication captures the average interaction density between qubits, offering insight into how strongly different parts of the circuit are coupled. Finally, liveliness provides a measure of qubit activity over time, indicating how long qubits remain engaged in the computation.

There are, of course, more features than the ones listed above that can be extracted from the circuit layout. As described in \appref{app:features}, we considered several additional such features, but excluded them from our feature vector. The reason for exclusion was either low importance for the final results or high
computational costs.



\subsection{Adding quantum hardware data}
\label{sec:AddingHardwareData}

In addition to structural circuit features, we also extract qubit-resolved features from the hardware that estimate gate failure probabilities arising from decoherence. These features are derived using the energy relaxation time $T_1$ and the dephasing time $T_2$ of each physical qubit, which are the dominant noise mechanisms in various quantum processors. We assume fixed gate durations for hardware-native operations, with $\tau_{1}$ and $\tau_{2}$ corresponding to typical single-qubit and two-qubit gate execution times, respectively. For each qubit $q_i$, we count:
\begin{itemize}
\item $N_i^{(1)}$: the number of single-qubit gates acting on $q_i$,
\item $N_i^{(2)}$: the number of two-qubit gates in which $q_i$ participates.
\end{itemize}

Consequently, the probability of decoherence-induced failure during a gate execution is approximated as~\cite{Nielsen2000, decayrate},
\begin{equation}
P_{\mathrm{fail}}(\tau) = \mleft(1 - e^{-\tau / T_1}\mright) + \mleft(1 - e^{-\tau / T_2}\mright),
\end{equation}
where relaxation and dephasing contributions are treated additively. For qubit $q_i$, the cumulative failure contribution from single-qubit gates is therefore defined as
\begin{equation}
F_i^{(1)} =
N_i^{(1)}
\mleft[
\mleft(1 - e^{-\tau_{1}/T_{1,i}}\mright)
+
\mleft(1 - e^{-\tau_{1}/T_{2,i}}\mright)
\mright],
\end{equation}
where $T_{1,i}$ and $T_{2,i}$ denote the relaxation and dephasing times, respectively, of qubit $q_i$. Similarly, the cumulative failure contribution from two-qubit gates involving qubit $q_i$ becomes
\begin{equation}
F_i^{(2)} =
N_i^{(2)}
\mleft[
\mleft(1 - e^{-\tau_{2}/T_{1,i}}\mright)
+
\mleft(1 - e^{-\tau_{2}/T_{2,i}}\mright)
\mright].
\end{equation}
This feature captures the increased exposure to decoherence due to longer two-qubit gate durations.

For qubits in the system that are not being used at all in the circuit, a large value is assigned for the failure rates,
\begin{equation}
F_i^{(1)} = F_i^{(2)} = 1000,
\end{equation}
indicating an effectively unusable or disconnected qubit. This value was chosen according to the maximum depth of the training circuits used.

In \appref{app:coherence}, we discuss other methods of encoding coherence times in the feature vector, and show that they do not yield as good performance for wPST prediction as the one presented in this section.



\subsection{Machine learning model}
\label{sec:MLModel}

The next part in our pipeline for estimating wPST and PST is to choose an accurate and efficient machine learning model. While neural-network models are often used for these prediction tasks, decision tree-based models have also been successfully used previously for FoM prediction~\cite{fompaper}. A decision tree~\cite{Breiman2017} is a supervised learning model that recursively partitions the feature space into disjoint regions in order to predict a target variable. Given a training dataset $\{(\mathbf{x}_i, y_i)\}_{i=1}^N$, where $\mathbf{x}_i \in \mathbb{R}^d$ denotes the feature vector and $y_i$ the corresponding target (FoM), the tree is constructed by iteratively selecting a feature and a split threshold that best separates the data according to a chosen impurity criterion. 
 
We selected different decision tree-based models and tested their performance and efficiency for the FoM prediction. These models included several standard techniques like random forest~\cite{Breiman2001}, bagging~\cite{Breiman1996}, etc., and modern methods like XGBoost~\cite{10.1145/2939672.2939785} and CatBoost~\cite{dorogush2018catboostgradientboostingcategorical}, etc. After evaluating the test results, which are shown in \appref{app:OtherModels}, we choose XGBoost as the most compatible for the FoM prediction, since it combined low training time and prediction time with low prediction errors. XGBoost is an optimized distributed gradient-boosting library designed to be highly efficient, flexible, and portable.

We optimized the hyperparameters for the models for different quantum devices using the RandomizedSearchCV module~\cite{scikitlearnRandomizedSearchCV} from Scikit--learn~\cite{scikit}. The final optimized hyperparameters are presented in \tabref{tab:xgb_hyperparams} in \appref{app:hyperparameters}.  


\subsection{Training data}

The training was done using randomly generated quantum circuits. Each circuit is defined on $n$ qubits and has a prescribed circuit depth $D$. At each layer of the circuit, a quantum gate is randomly selected and applied from a predefined gate set. The choice between single-qubit and two-qubit gates is governed by a tunable probability parameter $r \in [0,1]$. 

The single-qubit gate set consists of a mixture of Clifford, non-Clifford, and parameterized rotation gates:
\begin{equation}
\mathcal{G}_{1q} =
\mleft\{
\begin{aligned}
& R_X(\theta), R_Y(\theta), R_Z(\theta), \\
& U_1(\lambda), U_2(\phi,\lambda), U_3(\theta,\phi,\lambda), \\
& H, X, Y, Z, S, T
\end{aligned}
\mright\} ,
\end{equation}
where
\begin{align}
U_1(\lambda) &= \begin{pmatrix}1&0\\0&e^{i\lambda}\end{pmatrix}, 
\label{eq:U1}
\\
U_2(\phi,\lambda) &= \frac{1}{\sqrt{2}}\begin{pmatrix}1&-e^{i\lambda}\\e^{i\phi}&e^{i(\phi+\lambda)}\end{pmatrix} , \\
U_3(\theta,\phi,\lambda) &= \begin{pmatrix}\cos\mleft(\tfrac{\theta}{2}\mright)&-e^{i\lambda}\sin\mleft(\tfrac{\theta}{2}\mright)\\e^{i\phi}\sin\mleft(\tfrac{\theta}{2}\mright)&e^{i(\phi+\lambda)}\cos\mleft(\tfrac{\theta}{2}\mright)\end{pmatrix} .
\label{eq:U3}
\end{align}
For parameterized gates, all rotation angles are sampled independently from a uniform distribution over $[0,2\pi)$. Each single-qubit gate is applied to a qubit chosen uniformly at random from the set $\{0,\dots, n-1\}$.

The two-qubit gate set includes both entangling Clifford gates and controlled parameterized rotations:
\begin{equation}
\mathcal{G}_{2q} =
\mleft\{
\begin{aligned}
& \mathrm{CX}, \mathrm{CY}, \mathrm{CZ}, \mathrm{CH}, \mathrm{CR_X}(\theta), \\
& \mathrm{CR_Y}(\theta), \mathrm{CR_Z}(\theta), \mathrm{SWAP}
\end{aligned}
\mright\}.
\end{equation}
For controlled rotation gates, the rotation angle $\theta$ is sampled uniformly from $[0,2\pi)$. Each two-qubit gate acts on an unordered pair of distinct qubits chosen uniformly at random, thereby inducing non-local interactions across the register. 
The XGBoost model can be trained within a few seconds for a moderate data size (up to 100,000 circuits), and the predictions can be made in $\mathcal{O}(10^{-3})$ seconds. See \appref{app:OtherModels} for further details on training and testing time for this and other models.


\subsection{Performance metric}

Performance results for FoMs are often explained with the help of Pearson coefficients. The Pearson correlation coefficient quantifies the strength and direction of the linear relationship between two real-valued random variables $X$ and $Y$. It is defined as
\begin{equation}
\rho_{X,Y}
=
\frac{\mathrm{cov}(X,Y)}{\sigma_X \sigma_Y},
\end{equation}
where $\mathrm{cov}(X,Y)$ denotes the covariance between $X$ and $Y$, and
$\sigma_X$ and $\sigma_Y$ are the standard deviations of $X$ and $Y$, respectively.

Given $n$ paired observations $\{(x_i, y_i)\}_{i=1}^n$, the sample Pearson correlation coefficient is computed as
\begin{equation}
r
=
\frac{\sum_{i=1}^{n} (x_i - \bar{x})(y_i - \bar{y})}
{\sqrt{\sum_{i=1}^{n} (x_i - \bar{x})^2}
 \sqrt{\sum_{i=1}^{n} (y_i - \bar{y})^2}},
 \label{eq:pcorrelation}
\end{equation}
where $\bar{x}$ and $\bar{y}$ denote the sample means of $X$ and $Y$.

The Pearson coefficient satisfies $r \in [-1,1]$. A value of $r = 1$ indicates perfect positive linear correlation, $r = -1$ indicates perfect negative linear correlation, and $r = 0$ implies the absence of linear correlation. The performance of our machine learning models is judged on the value of the Pearson coefficient when the predicted and true FoMs are compared ($|r|=1$ implies a perfect result). Although nonlinear correlation metrics could potentially capture additional dependencies between different FoMs, such relationships are expected to include a linear or anti-linear component that is quantified by the Pearson correlation coefficient.


\subsection{Benchmark data set}

We tested all our models on random circuits and the  Munich Quantum Toolkit (MQT) benchmark dataset~\cite{2023mqtbench}, which is a standardized collection of quantum circuits designed to support systematic evaluation of quantum compilation, optimization, and execution performance across diverse algorithmic workloads. The dataset comprises circuits drawn from a wide range of application domains, including quantum chemistry, arithmetic, combinatorial optimization, simulation, and structured algorithmic benchmarks.



\section{Results and discussion}
\label{sec:results}

In this section, we present results for the PST and wPST machine learning prediction models for different quantum devices. Before testing on real quantum hardware, we tested the machine learning pipeline on a quantum simulator. Our goal is to evaluate how these models perform under different hardware constraints, noise characteristics, and connectivity topologies, and to assess their effectiveness as FoMs for quantum circuit compilation. At the end of this section, we also investigate how our new FoM, wPST, behaves in the presence of different types of noise on the quantum hardware.


\subsection{Testing on a quantum simulator}

We first tested the performance of the machine learning model on a quantum simulator, a 12-qubit device based on the first circular ring of the IBM Strasbourg quantum computer. For the quantum simulator, 1,200 random circuits for each number of qubits (14,400 circuits in total) with depth ranging from 1 to 100 (pre-transpilation) were used for training, and testing was done on 100 random circuits for each number of qubits (1,200 circuits in total) and 170 MQT benchmark circuits.

\begin{figure}
    \centering
    \includegraphics[width=\linewidth]{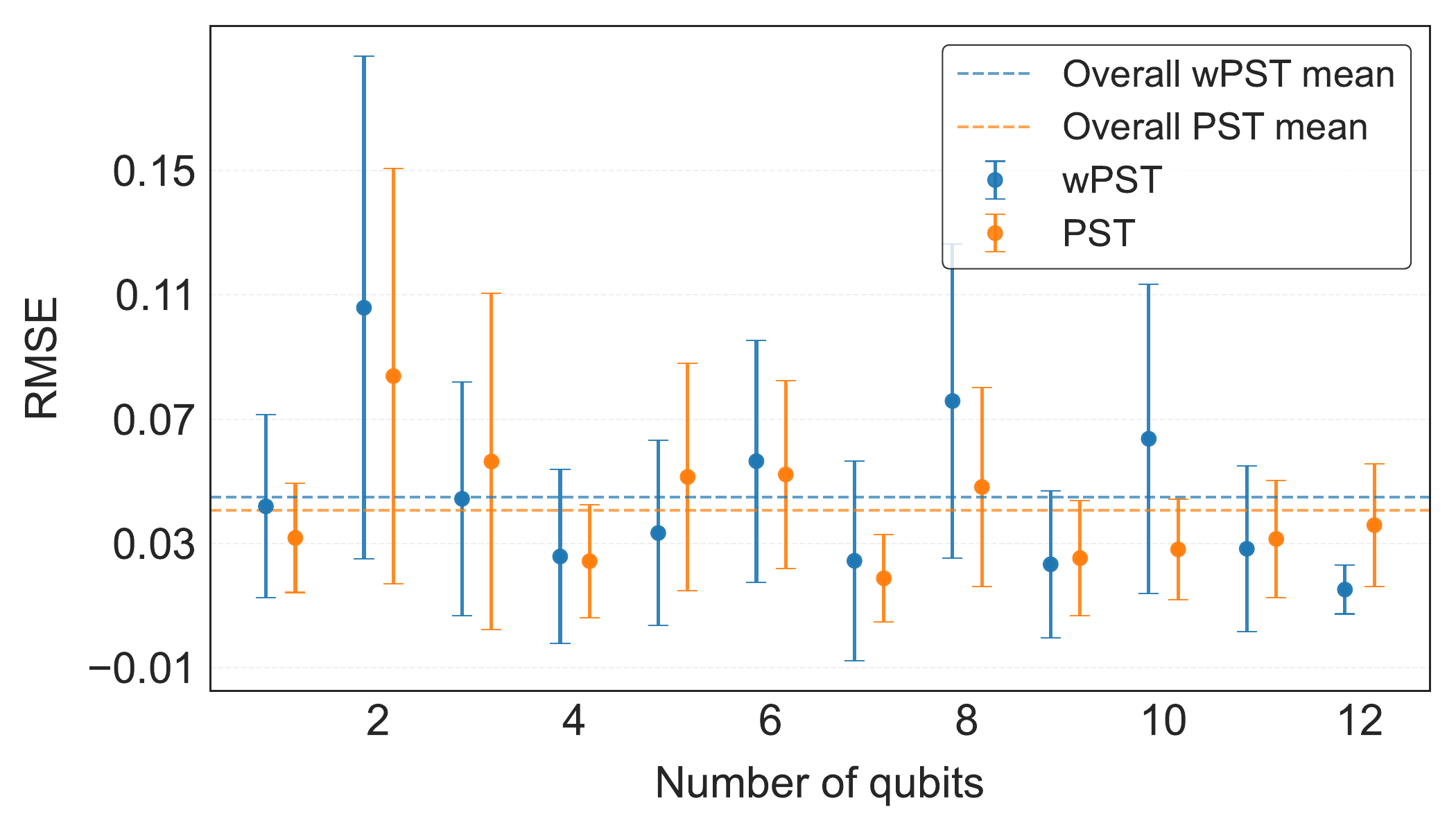}
    \caption{Root mean square error (RMSE) between the true and predicted PST and wPST for random circuits across qubits for the 12-qubit quantum simulator, with error bars showing $\pm 1$ standard deviation.}
    \label{fig:rmsesim}
\end{figure}

In \figref{fig:rmsesim}, we show the results from the testing on random circuits. Here, we plot the root mean square error (RMSE) between the PST and wPST predicted by our machine learning model and the true PST and wPST that we can obtain in the simulator. We observe that the RMSE stays quite constant for both the PST and wPST predictions as the number of qubits grows, which is promising for the scalability of the machine learning model.

\begin{table}
    \centering
    \caption{Pearson correlation coefficients for various metrics with respect to true PST for the 12-qubit quantum simulator.}
    \label{tab:pst_correlation}

\renewcommand{\arraystretch}{1.1}  
\setlength{\tabcolsep}{4pt}      
    
    \begin{tabular}{@{}lcc@{}}
        \toprule
        \textbf{Figure of merit} & \textbf{Random circuits} & \textbf{MQT benchmarks} \\
        \midrule
        Circuit depth & 0.790 & 0.585 \\
        \# Two-qubit gates & 0.810 & 0.520 \\
        \# Gates & 0.815 & 0.535 \\
        \textbf{Predicted PST} & \textbf{0.945} & \textbf{0.915} \\
        \bottomrule
    \end{tabular}
\end{table}

\begin{table}
    \centering
    \caption{Pearson correlation coefficients for various metrics with respect to true wPST for the 12-qubit quantum simulator.}
    \label{tab:wpst_correlation}

\renewcommand{\arraystretch}{1.1}   
\setlength{\tabcolsep}{4pt}          
    
    \begin{tabular}{@{}lcc@{}}
        \toprule
        \textbf{Figure of merit} & \textbf{Random circuits} & \textbf{MQT benchmarks} \\
        \midrule
        Circuit depth & 0.885 & 0.695 \\
        \# Two-qubit gates & 0.780 & 0.650 \\
        \# Gates & 0.805 & 0.650 \\
        \textbf{Predicted wPST} & \textbf{0.925} & \textbf{0.880} \\
        \bottomrule
    \end{tabular}
\end{table}

In Tables~\ref{tab:pst_correlation} and~\ref{tab:wpst_correlation}, we show the Pearson correlation coefficients of different FoMs, including both the circuit-based ones like circuit depth, number of gates, and number of two-qubit gates, and the predicted PST and wPST from our machine learning model, with the true PST (\tabref{tab:pst_correlation}) and wPST (\tabref{tab:wpst_correlation}) for both random and MQT circuit tests on the quantum simulator. These results demonstrate that our trained machine learning models clearly outperform individual circuit-level metrics in predictive capability for both PST and wPST. The traditional features of circuit depth, number of two-qubit gates, and total gate count only exhibit moderate correlation with the true performance, with values ranging between 0.78 and 0.89 for random circuits and between 0.52 and 0.70 for the MQT circuits. Notably, these correlations are consistently lower for the MQT circuits, indicating that simple metrics are insufficient to capture the more structured and heterogeneous behavior of benchmark workloads. In contrast, the predicted PST and wPST obtained from the trained model achieve significantly higher correlations across both datasets, reaching all the way to 0.945 and 0.925 for random circuits and maintaining a strong performance of 0.915 and 0.880 for the MQT benchmarks. The relatively small drop in correlation from random to MQT circuits suggests that the model generalizes well across different circuit classes, making it a robust and reliable estimator of circuit performance.


\subsection{Testing on quantum hardware}
\label{sec:TestingOnHardware}

Next, we extended the testing to actual quantum hardware, using four different IBM devices: IBM Torino, IBM Strasbourg, IBM Miami, and IBM Brussels. Since the hardware data has to be encoded into the feature vector, the layout of the IBM devices had to be fixed. Also, because results from extremely large circuits are highly affected by noise, we limited the execution of quantum circuits to a particular region of the quantum devices. The overall size was further limited to $21$ qubits (IBM Strasbourg), $21$ qubits (IBM Torino), $20$ qubits (IBM Miami), and $51$ qubits (IBM Brussels). The coupling maps of these devices, with the used qubits highlighted, are shown in \figref{fig:IBMdevices} of \appref{app:devices}.

Since the data available from the actual hardware is limited, a subset of the MQT benchmark dataset was used alongside the random circuits during training, and the model was tested against the other unseen subset of both random and MQT benchmark circuits. For IBM Strasbourg and IBM Torino, the models were trained using 2,100 random and 200 MQT benchmark circuits; the testing phase then evaluated these models on approximately 210 random and 200 MQT benchmark circuits. For IBM Miami, we utilized a smaller training set of 1,000 random and 200 MQT benchmark circuits, with testing performed on 200 samples from each set. Finally, IBM Brussels featured the largest dataset, with training involving 2,550 random circuits and 500 MQT benchmark circuits, while testing was conducted on 500 random and 454 MQT benchmark circuits.

In Tables~\ref{tab:pst_all_devices} and \ref{tab:wpst_all_devices}, we show the results in terms of the Pearson coefficient results across multiple quantum devices, in the same way as for the quantum simulator in Tables~\ref{tab:pst_correlation} and~\ref{tab:wpst_correlation}. The results in Tables~\ref{tab:pst_all_devices} and \ref{tab:wpst_all_devices} further reinforce the effectiveness and robustness of the proposed FoMs.

\begin{table*}[]
\centering
\caption{Pearson correlation coefficients with respect to true PST across four IBM devices, and their average.}
\label{tab:pst_all_devices}

\renewcommand{\arraystretch}{1.1}   
\setlength{\tabcolsep}{4pt}          

\begin{tabular}{@{}lcccccccccc@{}}
\toprule
\multirow{2}{*}{\textbf{Figure of merit}} & \multicolumn{2}{c}{\textbf{IBM Strasbourg}} & \multicolumn{2}{c}{\textbf{IBM Torino}} & \multicolumn{2}{c}{\textbf{IBM Miami}} & \multicolumn{2}{c}{\textbf{IBM Brussels}} & \multicolumn{2}{c}{\textbf{Average}} \\ \cmidrule(lr){2-3} \cmidrule(lr){4-5} \cmidrule(lr){6-7} \cmidrule(lr){8-9} 
 & \textbf{R} & \textbf{M} & \textbf{R} & \textbf{M} & \textbf{R} & \textbf{M} & \textbf{R} & \textbf{M}  \\ \midrule
Circuit depth           & 0.24 & 0.27 & 0.45 & 0.40 & 0.09 & 0.63 & 0.13 & 0.27 & & 0.31 \\
\# Two-qubit gates      & 0.43 & 0.45 & 0.58 & 0.35 & 0.56 & 0.40 & 0.28 & 0.12 & & 0.40 \\ 
\# Gates         & 0.41 & 0.32 & 0.56 & 0.37 & 0.59 & 0.48 & 0.25 & 0.14 & & 0.39 \\
\textbf{Predicted PST}  & \textbf{0.97} & \textbf{0.92} & \textbf{0.96} & \textbf{0.97} & \textbf{0.95} & \textbf{0.93} & \textbf{0.98} & \textbf{0.91} & & \textbf{0.95} \\ \bottomrule
\end{tabular}
\end{table*}

\begin{table*}[]
\centering
\caption{Pearson correlation coefficients with respect to true wPST across four IBM devices, and their average.}
\label{tab:wpst_all_devices}

\renewcommand{\arraystretch}{1.1}   
\setlength{\tabcolsep}{4pt}          

\begin{tabular}{@{}lcccccccccc@{}}
\toprule
\multirow{2}{*}{\textbf{Figure of merit}} & \multicolumn{2}{c}{\textbf{IBM Strasbourg}} & \multicolumn{2}{c}{\textbf{IBM Torino}} & \multicolumn{2}{c}{\textbf{IBM Miami}} & \multicolumn{2}{c}{\textbf{IBM Brussels}} & \multicolumn{2}{c}{\textbf{Average}} \\ \cmidrule(lr){2-3} \cmidrule(lr){4-5} \cmidrule(lr){6-7} \cmidrule(lr){8-9} 
 & \textbf{R} & \textbf{M} & \textbf{R} & \textbf{M} & \textbf{R} & \textbf{M} & \textbf{R} & \textbf{M}\\ \midrule
Circuit depth           & 0.57 & 0.44 & 0.67 & 0.58 & 0.69 & 0.68 & 0.44 & 0.14 &  & 0.53 \\
\# Two-qubit gates     & 0.65 & 0.26 & 0.52 & 0.54 & 0.66 & 0.56 & 0.37 & 0.21 &  & 0.47 \\
\# Gates         & 0.61 & 0.32 & 0.56 & 0.55 & 0.73 & 0.58 & 0.39 & 0.18 &  & 0.49 \\
\textbf{Predicted wPST} & \textbf{0.93} & \textbf{0.82} & \textbf{0.93} & \textbf{0.94} & \textbf{0.92} & \textbf{0.81} & \textbf{0.92} & \textbf{0.77} &  & \textbf{0.89} \\ \bottomrule
\end{tabular}
\end{table*}

In \tabref{tab:pst_all_devices}, we see that for true PST, traditional circuit metrics such as depth, total gate count, and number of two-qubit gates exhibit relatively weak and highly variable correlations across devices, with average values below or at $0.40$. Moreover, their performance is inconsistent between random (R) and MQT benchmark (M) circuits, highlighting their limited ability to generalize across different circuit structures and hardware characteristics. In stark contrast, the predicted PST achieves consistently high correlations across all devices, exceeding $0.95$ on average and maintaining strong performance for both random and benchmark circuits.

A similar trend is observed for wPST in \tabref{tab:wpst_all_devices}, where conventional features provide moderate correlations with an average around 0.5, while the predicted wPST significantly outperforms them with an average correlation of 0.89. Importantly, the model demonstrates stable behavior across diverse devices, with some degradation for benchmark circuits compared to random circuits.
\begin{figure*}
        \centering
        \includegraphics[width=\textwidth]{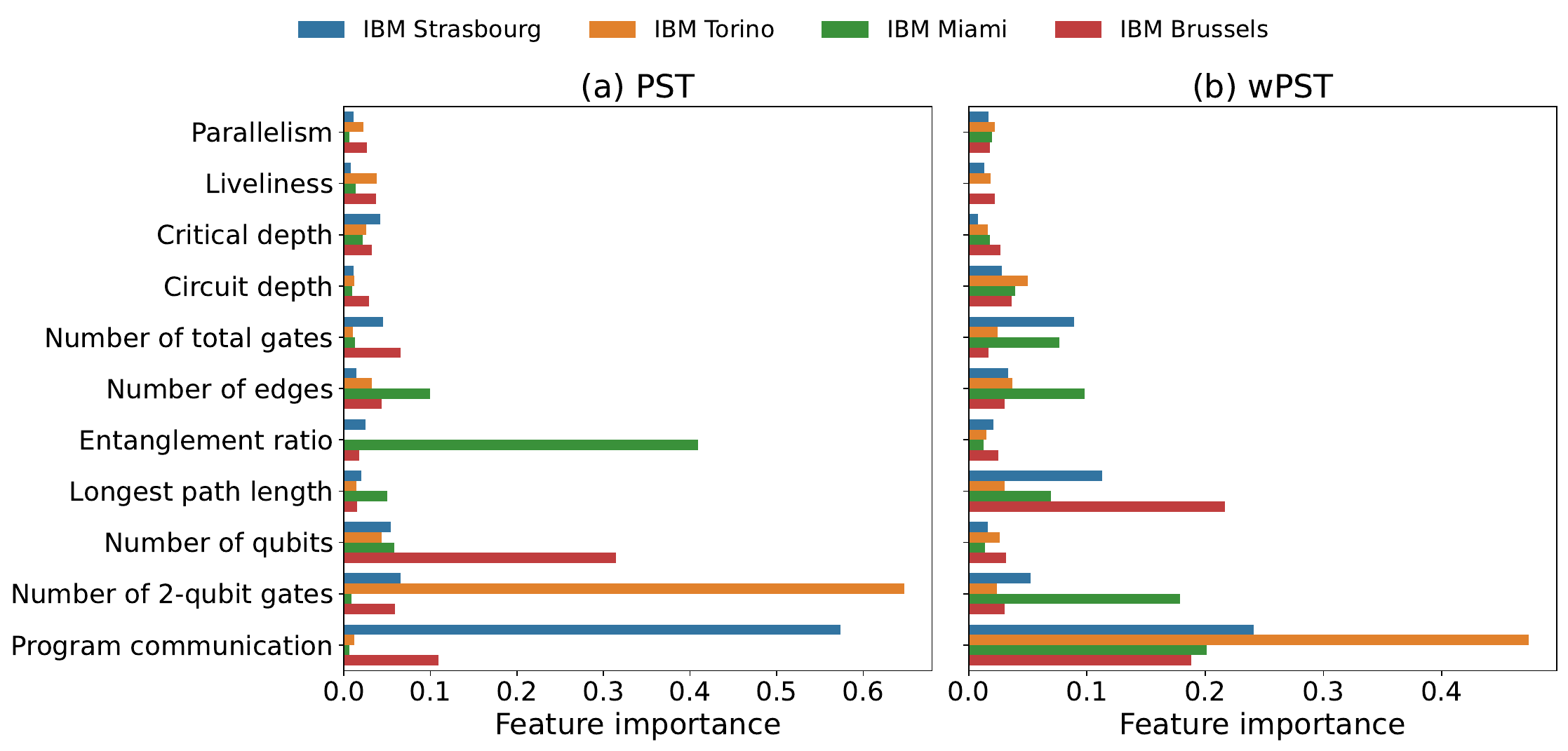} 
        \caption{Importance of different features across different quantum devices in the (a) PST and (b) wPST prediction models.}
        \label{fig:FeatureImportance}
\end{figure*}

We further explore the strong correlations that our machine learning models achieve for both PST and wPST, by determining the feature importance of the models, as shown in \figref{fig:FeatureImportance}. The figure shows that both models are mainly making use of the program communication and the number of two-qubit gates, with significant contributions from the entanglement ratio and the circuit depth, depending on the device. This shows that our proposed FoMs capture not only the physical attributes like depth and the amount of entangling gates very well, but also the qubit interactions. This makes for a very well-rounded FoM that can go beyond the commonly used ones.

\begin{table}
\centering
\caption{Prediction performance in simulations and across different quantum devices. Each device is evaluated on a set of random quantum circuits using the $R^2$ and RMSE metrics for both PST and wPST.}
\label{tab:device_performance}

\renewcommand{\arraystretch}{1.1}   
\setlength{\tabcolsep}{4pt}          

\begin{tabular}{lcccc}
\toprule
\textbf{Device} 
& \multicolumn{2}{c}{\textbf{PST}} 
& \multicolumn{2}{c}{\textbf{wPST}} \\
\cmidrule(lr){2-3} \cmidrule(lr){4-5}
& \textbf{$R^2$} & \textbf{RMSE} 
& \textbf{$R^2$} & \textbf{RMSE} \\
\midrule
Simulator        & 0.9655 & 0.0411 & 0.9846 & 0.0427 \\
IBM Strasbourg   & 0.9126 & 0.0802 & 0.8448 & 0.0771 \\
IBM Torino       & 0.9602 & 0.0667 & 0.8772 & 0.0599 \\
IBM Miami        & 0.8461 & 0.0465 & 0.8923 & 0.0584 \\
IBM Brussels     & 0.8999 & 0.1066 & 0.8197 & 0.0863 \\
\bottomrule
\end{tabular}
\end{table}

For completeness, we also present, in Table~\ref{tab:device_performance}, the RMSE and R2 scores across the simulator and different quantum devices for the machine learning models, tested on random quantum circuits. We note that we obtain the lowest errors on the simulator, the performance is also good for the actual hardware.
 

\subsection{Response of wPST to different noise channels}
\label{sec:noise}
For understanding the behavior and realistic performance of wPST, it is helpful to know how dominant noise channels~\cite{Gaitan2018} influence it. In this section, we examine the response of wPST to several canonical noise channels and derive the corresponding reference behaviors. These results provide a baseline for interpreting wPST values and distinguishing the effects of different noise processes on circuit performance.


\paragraph{Bit-flip noise.}

The most direct noise channel for the mirror framework is the independent bit-flip noise, 
\begin{equation}
    \mathcal{E}(\rho) = (1-\epsilon)\rho + \epsilon X\rho X.
\end{equation}
When applied to each qubit at readout, an $X$ error deterministically increments the Hamming weight of the measured string. If the flips instead occur mid-circuit, subsequent gates can convert them into phase errors that leave the computational-basis probabilities---and hence the wPST---unchanged, just as in the dephasing case below; the expressions here correspond to the limit in which all flips survive to readout. The number of correct bits is then binomially distributed, $k \sim \mathrm{Bin}(n, 1-\epsilon)$, giving the exact results $\mathrm{PST} = (1-\epsilon)^n$ and $\mathrm{wPST} = 1-\epsilon$. The former decays exponentially in $n$, while the latter is independent of $n$ (see \appref{app:bitflip} for derivations). Moreover, for $\epsilon < 1/2$, concentration of measure implies that the thresholded and unthresholded wPST coincide up to corrections exponentially small in $n$, so that in the large-$n$ limit
\begin{equation}\mathrm{wPST}_{w_{\rm th}}\xrightarrow{n \to \infty}
\begin{cases}
1-\epsilon & \text{if } w_{\rm th} \le 1-\epsilon , \\
0 & \text{if } w_{\rm th} > 1-\epsilon .
\end{cases}
\end{equation}
This provides a principled justification for the choice $w_{\rm th}=0.5$ in \eqref{eq:wpst}: the threshold is asymptotically inactive whenever the per-qubit error rate is better than random ($\epsilon < 1/2$), and only activates in the majority-incorrect regime. 


\paragraph{Depolarizing noise.}

Depolarizing noise sets the reference for a maximally scrambled circuit. 
Strong depolarizing noise [\eqref{eq:DepolarizingNoiseDef}] drives the qubits toward the maximally mixed state, in which every bit string occurs with probability $2^{-n}$. Evaluating \eqref{eq:wpsthamming} for this distribution yields the depolarizing reference
\begin{equation}
F(n; w_{\rm th}) = \frac{1}{2^n} \sum_{k=k_{\rm th}}^{n} \binom{n}{k} \frac{k}{n} ,
\label{eq:depolfloor}
\end{equation}
with
\begin{equation}
\begin{aligned}
F(n;0)&=\frac{1}{2},\\
F(n;1)&=2^{-n}=\mathrm{PST},\\
\lim_{n\rightarrow\infty}F(n;0.5)&=\frac{1}{4}.
\end{aligned}
\end{equation}
A fully scrambled circuit therefore does \emph{not} score $\mathrm{wPST} = 0$, but somewhere near $0.25$ at our threshold; a plateau of measured wPST values near this level should be read as the noise floor rather than as residual signal. 

Under the assumption that the accumulated noise acts as a single global depolarizing channel on the pre-measurement state, the measured output \emph{distribution} becomes a mixture of the ideal and uniform distributions,
\begin{equation}
    p(x) = (1-\lambda)\, p_{\rm ideal}(x) + \lambda\, 2^{-n},
\end{equation}
where $p(x)$ is the probability of measuring bit string $x$. Since wPST [\eqref{eq:wpst}] is linear  in these probabilities, we obtain 
\begin{equation}
    \mathrm{wPST} = (1-\lambda) + \lambda F \in [F, 1].
\end{equation}
Thus, $F$ acts as a lower bound within this noise model. We stress, however, that $F$ is a typical-noise reference, not a universal bound: bit-flip noise with $\epsilon > 1/2$ yields $\mathrm{wPST}_0 = 1-\epsilon < 1/2 = F(n;0)$, and coherent errors (see below) can concentrate probability mass in a specific incorrect region of Hamming space and land below the floor. The only unconditional lower bound is $\mathrm{wPST} \ge 0$.


\paragraph{Dephasing noise.}

The phase-flip noise, 
\begin{equation}
    \mathcal{E}(\rho) = (1-p)\rho + p Z\rho Z
\end{equation}
commutes with computational-basis measurement and thus does not flip bits directly. It can degrade the wPST only insofar as subsequent gates convert phase errors into bit errors. Its effect is therefore circuit-dependent, and no closed form analogous to the above expressions exists.


\paragraph{Amplitude damping.}

\begin{figure*}
        \centering
        \includegraphics[width=0.75\linewidth]{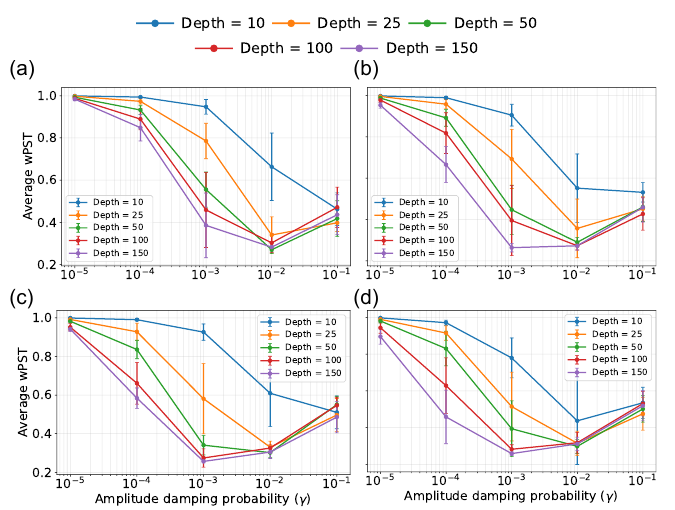} 
        \caption{Impact of different damping rates $\gamma$ on average wPST of circuits of different depths (lines of different colors) for systems of size (a) four, (b) six, (c) eight, and (d) ten qubits. The simulations were carried out on a ten-qubit simulator based on the IBM Miami device; 20 circuits were used for average results over each depth. The error bars show $\pm 1$ standard deviation.}
        \label{fig:damping}
\end{figure*}

Like dephasing, relaxation occurring mid-circuit has a circuit-dependent effect: a decay on a qubit in $\ket{1}$ acts as a bit flip that subsequent gates may propagate, cancel, or convert into a phase error. What distinguishes amplitude damping is its asymmetry---it only drives $\ket{1}\to\ket{0}$---so decay near the end of the circuit pushes measurement outcomes \emph{toward} the all-zero reference state.
Given the gate time $t$ and a qubit's energy relaxation time $T_1$, the amplitude-damping probability can be defined as 
\begin{equation}
    \gamma = 1 - e^{-t / T_1} .
\end{equation}
This asymmetric bias inflates both PST and wPST, as seen in \figref{fig:damping}, and is the failure mode of any benchmark referenced to $\ket{0^n}$, which wPST inherits from PST. For realistic damping rates ($\gamma \lesssim 10^{-3}$), wPST can nonetheless be reliably used to benchmark circuits of variable size and depth, as shown in \figref{fig:damping}. Only for extreme damping ($ \gamma > 10^{-3}$) does one see a significant inflation in wPST values. In the realistic regime, circuits up to depth $100$ (pre-transpilation) can be quantified by wPST, and that is the range chosen for all the training data used in this work.
There are ways to counter the damping problem, as this bias could be suppressed by using randomized compiling (Pauli twirling)~\cite{twirl1, twirl2, twirl3}, which tailors the channel into stochastic Pauli noise, but we leave this to future work.


\paragraph{Readout errors.}

Readout errors act only on the final measurement layer and are typically asymmetric, $p(1|0) \ne p(0|1)$. For PST, readout errors act multiplicatively across qubits, contributing an additional factor of roughly $[1-p(1|0)]^n$ to the survival probability and supporting the existing exponential decay. For wPST, by \eqref{eq:wpstmarginals}, a single-qubit readout flip costs only $1/n$, so readout errors enter additively rather than multiplicatively. This makes wPST considerably more robust to readout errors at scale than PST.


\paragraph{Coherent errors.}

Finally, systematic over- or under-rotations are unitary and accumulate constructively across a circuit rather than averaging out stochastically. Unlike the Pauli channels above, they can steer the output toward specific incorrect bit strings (possibly with high Hamming weight), which is one mechanism by which measured wPST values can fall below the depolarizing reference $F$.


\paragraph{Total effects of errors.}

In summary, wPST degrades linearly with the per-qubit error rate under stochastic bit-flip-like noise at readout, saturates at the floor in \eqref{eq:depolfloor} under strong depolarization, is inflated by strong amplitude damping, and is comparatively insensitive to phase and readout errors. Together, these results demonstrate that wPST provides a more informative characterization of the circuit execution than PST, as it captures the gradual accumulation of local errors while remaining robust to system-size scaling. Consequently, wPST provides a more representative measure of circuit reliability in larger quantum systems, where exact output recovery becomes increasingly stringent and less informative.



\section{Two-step prediction process: Figure of merit for quantum compilers}
\label{sec:2step}

The results presented in \secref{sec:results} were for circuits that already had been transpiled for a particular quantum device. However, for an FoM to be useful in quantum compilation, it should be able to figure out a way to quantify non-transpiled quantum circuits that have to be evaluated during the compilation process. In this section, we show how to adapt our machine learning pipeline to handle such circuits.


\subsection{Details of the two-step process}
\label{sec:2StepDetails}

Quantum circuit compilation is the process of transforming an abstract, hardware-agnostic quantum algorithm into an executable circuit on a specific quantum device. This transformation typically involves several stages, including gate decomposition, qubit mapping, routing, and optimization~\cite{compil1, compil2}. High-level quantum algorithms are often expressed using an idealized, universal gate set and assume full qubit connectivity. However, real quantum devices have a limited native gate set and exhibit limited connectivity, as described by a coupling map. Consequently, the compiler must decompose unsupported gates into sequences of native operations and insert additional operations, such as SWAP gates, to satisfy connectivity constraints.

To connect the wPST machine learning predictor to a quantum compiler, we devise a two-step prediction process, where the first step consists of predicting the feature vector of the transpiled quantum circuit to be executed on a given quantum hardware. This feature vector is then used to obtain a final prediction for the FoM in the second step. This process allows us to approximately introduce the impact of both the coupling maps and the quantum hardware into the model.

To encode hardware connectivity information, we extract degree-based features from the device coupling map. These features capture the local connectivity of each physical qubit and provide a compact representation of hardware topology constraints. The coupling map of a quantum device is modeled as a graph
\begin{equation}
G_{\mathrm{c}} = (\mathcal{V}, \mathcal{E}),
\end{equation}
where
\begin{itemize}
\item $\mathcal{V} = \{q_1, q_2, \ldots, q_n\}$ represents the set of physical qubits and
\item $\mathcal{E} \subseteq \mathcal{V} \times \mathcal{V}$ represents allowed two-qubit interactions.
\end{itemize}

Although the physical coupling map may be directed, we treat it as an undirected graph when computing degree-based features, as bidirectional interaction capability is typically enforced through calibration and transpilation. 

The degree of a qubit $q_i$ is defined as
\begin{equation}
d_i = |\{ q_j \in \mathcal{V} \mid (q_i, q_j) \in \mathcal{E} \ \text{or} \ (q_j, q_i) \in \mathcal{E} \}|.
\end{equation}
This quantity measures the number of distinct qubits with which $q_i$ can directly participate in a two-qubit gate. The coupling-degree vector is defined as
\begin{equation}
\mathbf{d} =
\mleft[
d_1, d_2, \ldots, d_n
\mright] \in \mathbb{R}^{n},
\end{equation}
where $n$ is the total number of physical qubits. Each entry in the coupling-degree vector is computed by iterating over the coupling list and incrementing the degree count of both endpoints of each edge:
\begin{equation}
d_i \leftarrow d_i + 1, \quad
d_j \leftarrow d_j + 1,
\quad \forall (q_i, q_j) \in \mathcal{E}.
\end{equation}

We also tried other methods of coupling-map encoding. Those methods are discussed further in \appref{app:coupling}. 

\begin{figure*}
    \centering
        \includegraphics[width=\textwidth]{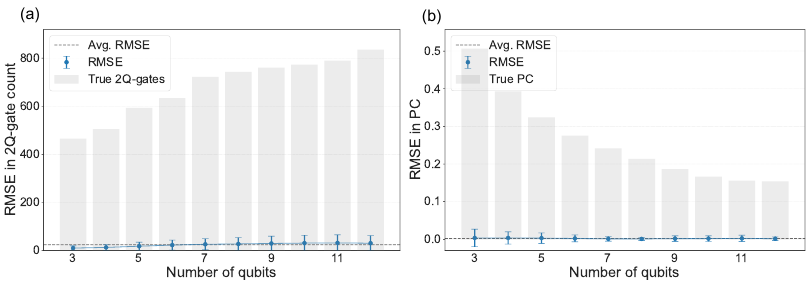} 
        \caption{Root mean square error in prediction of (a) the number of two-qubit gates and (b) programme communication (PC) for a 12-qubit system on a simulator, with error bars showing $\pm 1$ standard deviation. The dashed lines show the average RMSE across different numbers of qubits and the shaded bars show the average true counts.}
        \label{fig:rmse2}
\end{figure*}

Using the coupling-map data, we can encode the extended feature vector of the non-compiled circuit and predict the feature vector of the compiled circuit. Generally, a multi-output decision tree chooses splits to reduce the total variance across all outputs. As the number of outputs increases, it becomes harder to find splits that improve all outputs simultaneously, leading to weaker splits. However, because quantum circuit features are correlated and structured, multi-output regression trees can still give accurate predictions despite the multi-output setup. In a general case, it might be beneficial to predict only a couple of important features (e.g., the number of two-qubit gates and program communication, as shown in \figref{fig:rmse2} and as examples for other coupling-map encodings in ~\figref{fig:rmseApp} in \appref{app:coupling}) and then update the other parameters accordingly. Otherwise, for a particular device where an ample amount of data is available, one can predict the entire feature vector.


\subsection{Performance of the two-step process}
\label{sec:2StepPerformance}

\begin{figure}
    \centering
    \includegraphics[width=\linewidth]{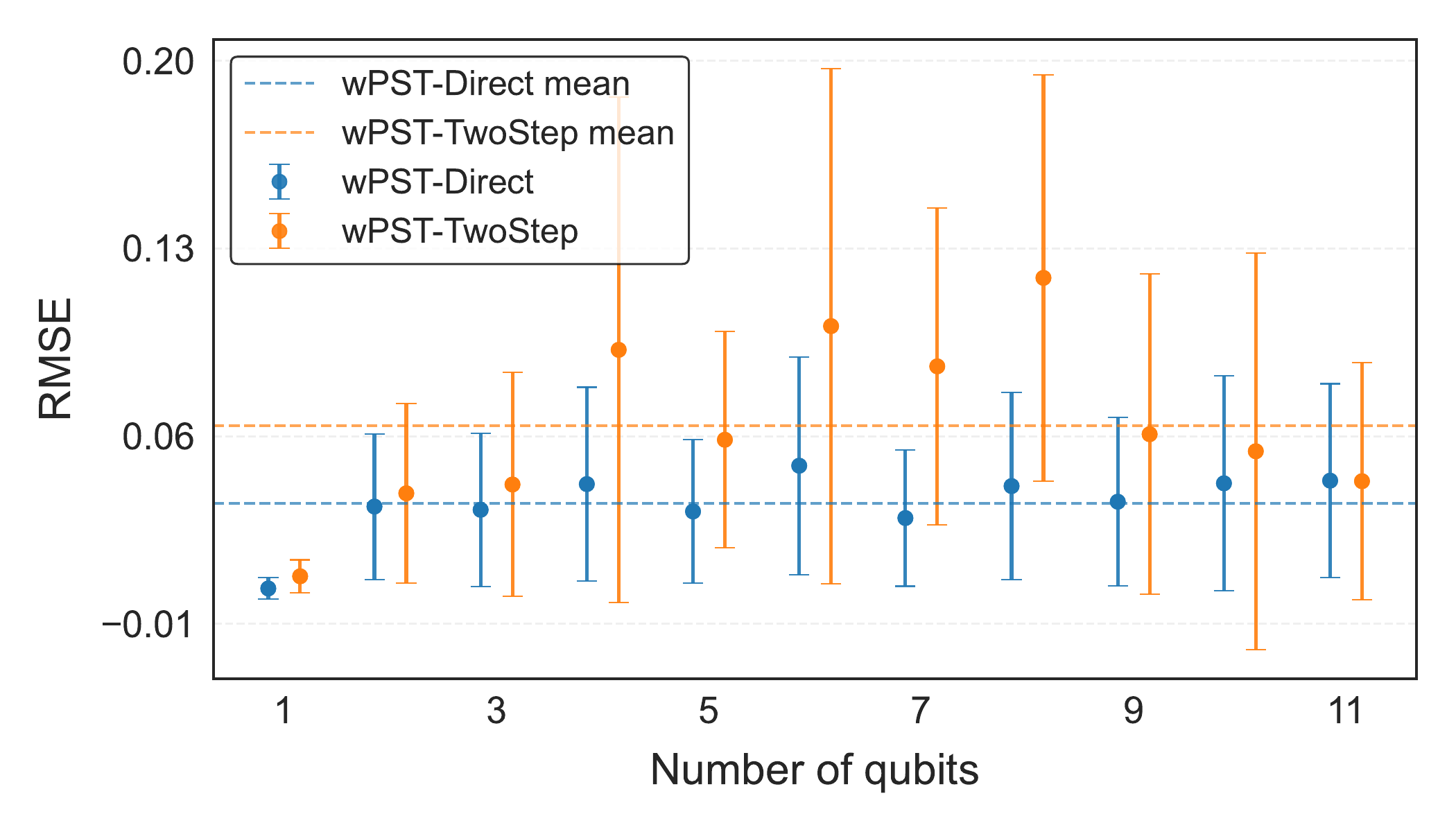}
    \caption{Root mean squared error for predicted wPST using direct and two-step schemes in the IBM Torino device, with dashed lines showing the overall mean and the error bars showing $\pm 1$ standard deviation.}
    \label{fig:rmsetwostep}
\end{figure}

In \figref{fig:rmsetwostep}, we show results from benchmarking the performance of the two-step predictor against the direct wPST predictor on the IBM Torino device. Just like earlier, 1,200 random circuits were used for training while 240 random circuits were used for testing. As expected, the error rates are higher when the two-step process is employed, but the two-step process is still quite accurate (average RMSE $\sim$0.06) across the range of qubit numbers. However, before the two-step process can be employed in quantum compilers, additional research is required into which features to choose and how the quantum compilers will perform using this process.


\section{Conclusion and outlook}
\label{sec:conclusion}

\begin{figure}
    \centering
    \includegraphics[width=\linewidth]{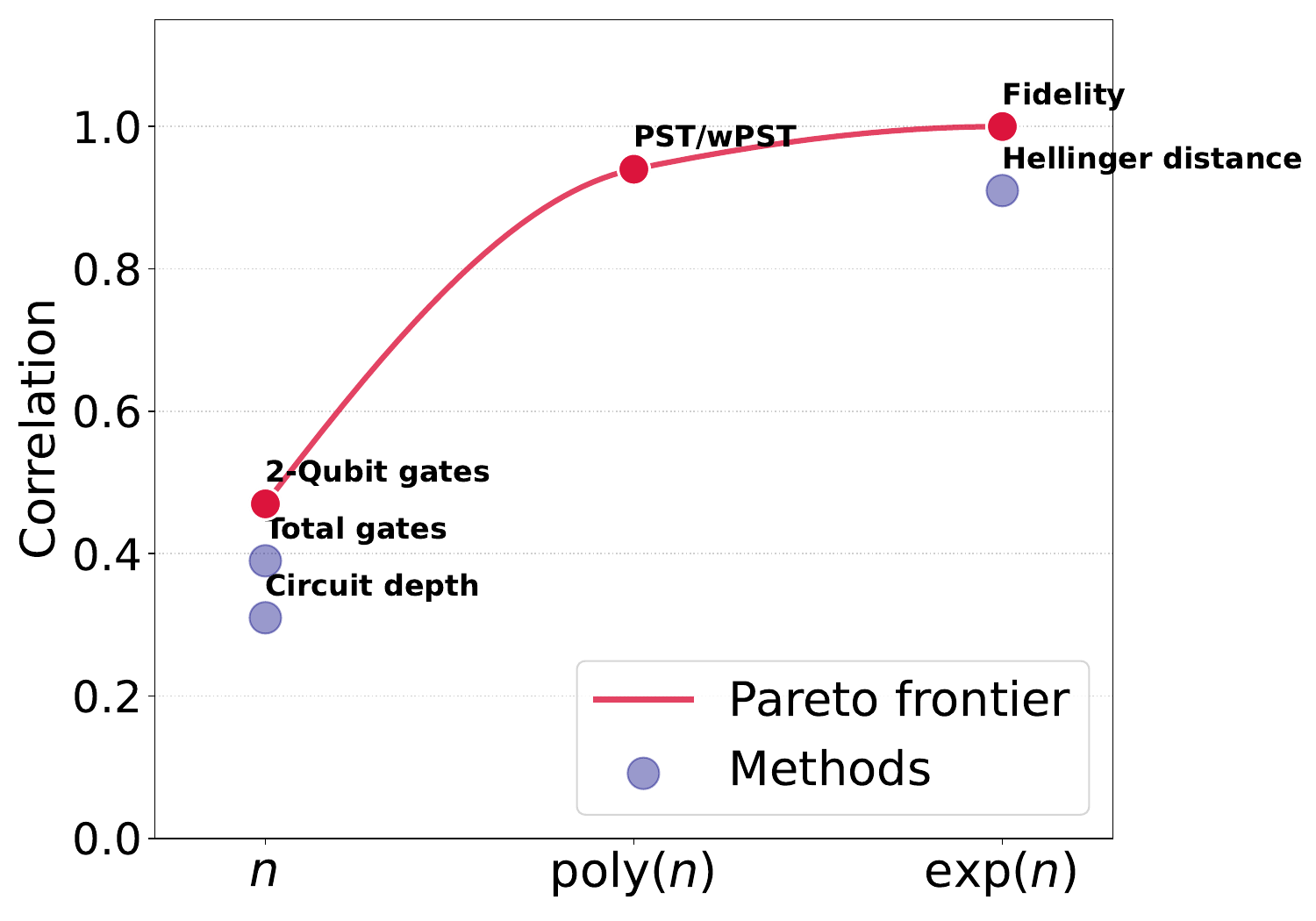}
    \caption{Illustration of the trade-off between model accuracy (here in terms of correlation with either algorithmic fidelity, distance, or true PST/wPST) and computational complexity for different FoMs for quantum circuit compilation. The Pareto frontier and the correlation values for different FoMs are estimates mostly based on numerical simulations or experimental tests in this article (see Tables~\ref{tab:pst_correlation}--\ref{tab:wpst_all_devices}), not rigorously proven general results. The point for PST/wPST refers to our estimate using a machine learning model. The result for the Hellinger distance is taken from Ref.~\cite{fompaper}.}
    \label{fig:pareto}
\end{figure}

In this article, we examined figures of merit (FoMs) for evaluating quantum circuit quality (how well a quantum algorithm is executed) in the NISQ era, where limited hardware capabilities fundamentally constrain the execution of quantum circuits. 
Our contribution consists of three main results: (i) the introduction of a weighted probability of successful trials (wPST) as a FoM with good properties, (ii) the construction and benchmarking of a machine learning model for estimating wPST with low computational cost, and (iii) the architecture of a two-step prediction process for wPST, applicable in quantum compilation where a FoM of non-transpiled circuits must be estimated.

Ideally, FoMs should be able to evaluate circuit quality both accurately and quickly, but in practice, there is a trade-off between these two goals. Figure~\ref{fig:pareto} illustrates this trade-off, showing how different FoMs occupy distinct regions along the Pareto frontier. Simple structural features such as circuit depth, total gate count, and number of two-qubit gates are quick to calculate, but their correlation with true performance (algorithmic fidelity, distance, or true PST/wPST) remains limited, indicating that they fail to capture the full complexity of circuit behavior. On the opposite end, metrics grounded in actually running or simulating the quantum circuit, e.g., Hellinger distance and fidelity, incur significantly higher computational overhead. 

In this context, the predicted probability of successful trials (PST) and its weighted version, wPST, which we proposed here, emerge as an effective compromise, achieving high correlation values close to the true metrics while maintaining polynomial complexity with the system size (number of qubits). This positioning along the Pareto frontier in \figref{fig:pareto} highlights their ability to balance accuracy and efficiency, effectively identifying the ``sweet spot" where predictive performance is maximized without incurring the prohibitive costs associated with full quantum evaluation. As a result, PST and wPST provide a practical and scalable alternative for performance estimation in quantum compilation workflows. In particular, our proposed wPST avoids a particular drawback of PST and fidelity---the inability to distinguish single- and multiple-bit-flip errors. Indeed, we have shown that this difference in turn makes wPST better behaved with respect to noise than PST in several ways, and that the number of circuit runs required for measuring wPST scales better with system size than for measuring PST.

The key to lower computational complexity, and thereby enable the use of these FoMs for quantum compilation at scale, was that we developed machine learning models that accurately predict PST and wPST by incorporating both circuit-level features and hardware-specific information. This approach enables fast and reliable estimation of circuit quality without the need for repeated quantum executions, making it well suited for use in quantum compilation and optimization workflows. The machine learning-predicted FoMs consistently outperformed commonly used FoMs based on circuit features (e.g., the number of gates), yielding, on average, a nearly \SI{50}{\percent} improvement in correlation with true PST and wPST, in both numerical simulations and in tests on several IBM quantum processors.

To handle non-transpiled circuits, which must be evaluated partway through quantum compilation, we extended the machine-learning framework for wPST by introducing a two-step prediction pipeline. In that pipeline, we first estimate the additional quantum resources that hardware connectivity constraints will add to the non-transpiled circuit. Then, in the second step, we use this estimate together with coherence data for the quantum processor to predict the final wPST. Since the two-step prediction is fully based on trained machine learning models, the computational cost remains low, such that it continues to be possible to use this FoM directly in quantum compilers.

While our results constitute a significant step forward for machine learning models for FoM prediction, there are many challenges in the bigger picture of quantum compilation. A first task for follow-up work would be to integrate our proposed FoM and its prediction by machine learning with a quantum compiler, and compare the resulting compilation performance with that achieved for other FoMs. There is also the fundamental question of `double' depth in wPST calculations. Since wPST calculations involve measurements of circuits double the size of the original circuits, there will always be the question of true representation, that is, whether PST-based methods are truly highlighting the effect of noise on the original quantum circuits. More generally, further exploration of the Pareto frontier sketched in \figref{fig:pareto} is called for, both in terms of which FoM navigates the trade-off between accuracy and computational complexity best (that may differ depending on the application) and in terms of how machine learning can help estimate FoMs to improve the trade-off.


\begin{acknowledgments} 

HS thanks Caspar Groiseau and Jingjun Zhu for useful discussions during the initial phases of the project.
This work was supported by the Swedish Foundation for Strategic Research (grant number FUS21-0063) and the Knut and Alice Wallenberg Foundation through the Wallenberg Centre for Quantum Technology (WACQT). 
AFK is also supported by the Swedish Foundation for Strategic Research (grant number FFL21-0279) and the Horizon Europe programme HORIZON-CL4-2022-QUANTUM-01-SGA via the project 101113946 OpenSuperQPlus100.

\end{acknowledgments} 


\appendix


\section{Quantum hardware used in testing}
\label{app:devices}

\begin{figure*}
    \centering
    \includegraphics[width=\linewidth]{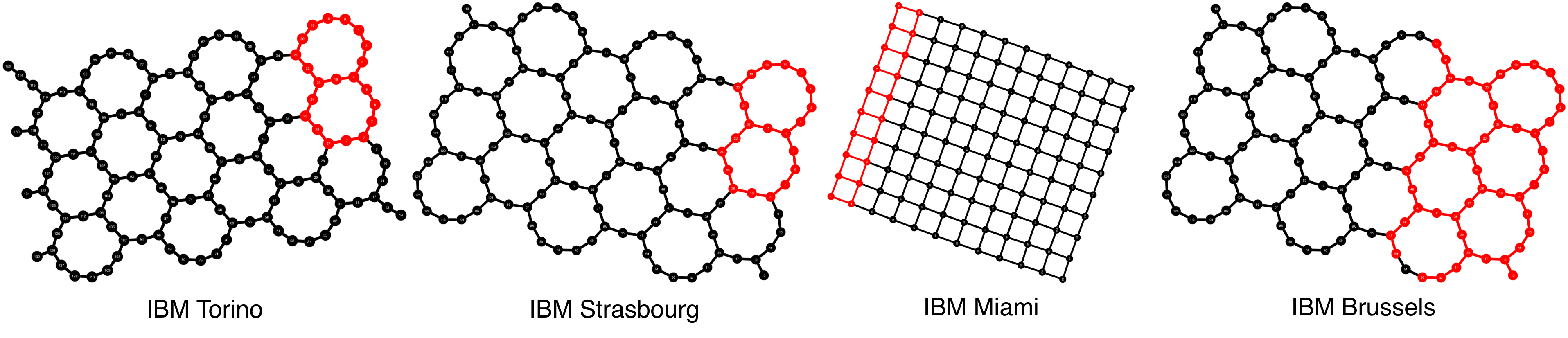}
    \caption{Layouts of the four IBM quantum processors we used for training and testing in this article. The orange nodes mark the subsets of qubits we used on each device.}
    \label{fig:IBMdevices}
\end{figure*}

The results presented in Secs.~\ref{sec:TestingOnHardware} and \ref{sec:2StepPerformance} from training and testing on quantum hardware were obtained using four quantum processors from IBM. In \figref{fig:IBMdevices}, we show the coupling maps of these four devices, highlighting the subset of selected qubits on each device that was used for training and testing.


\section{XGBoost hyperparameters}
\label{app:hyperparameters}


\begin{table*}
\centering
\caption{Optimized XGBoost hyperparameters obtained for the wPST prediction model for different systems.}
\label{tab:xgb_hyperparams}
\renewcommand{\arraystretch}{1.1}   
\setlength{\tabcolsep}{4pt}          
\begin{tabular}{lccccc}
\toprule
\textbf{Hyperparameter} 
& \textbf{Simulator} 
& \textbf{Torino} 
& \textbf{Strasbourg} 
& \textbf{Miami} 
& \textbf{Brussels} \\
\midrule
colsample\_bytree & 0.6704 & 0.7125 & 0.9838 & 0.6804 & 0.7217 \\
gamma             & 0.0904 & 0.1520 & 0.0197 & 0.0652 & 0.1721 \\
learning\_rate    & 0.1088 & 0.0192 & 0.0510 & 0.1141 & 0.2086 \\
max\_depth        & 6      & 9      & 6      & 6      & 9      \\
min\_child\_weight& 8      & 2      & 2      & 6      & 1      \\
reg\_alpha        & 0.6496 & 0.9987 & 0.6269 & 0.1167 & 0.7053 \\
reg\_lambda       & 2.1984 & 0.6615 & 2.3191 & 0.6343 & 0.6306 \\
subsample         & 0.8630 & 0.9940 & 0.6065 & 0.6414 & 0.7157 \\
\bottomrule
\end{tabular}
\end{table*}

As described in \secref{sec:MLModel}, we optimized the hyperparameters of XGBoost for wPST prediction in several systems. The hyperparameters resulting from that optimization, and used to obtain the results in Secs.~\ref{sec:results} and \ref{sec:2step}, are shown in Table~\ref{tab:xgb_hyperparams}.


\section{Other circuit features}
\label{app:features}

In \secref{sec:EncodingCircuits}, we listed features derived from the DAG representation of a quantum circuit or from its gate-level description, which were included in the feature vector used in our machine learning models. Here, for completeness, we discuss a number of features that were excluded from the feature vector, mainly due to either low importance for the final results or high computational costs.

Similar to keeping a count of two-qubit gates, one can also track the number of single-qubit U1, U2, and U3 gates [see Eqs.~(\ref{eq:U1})--(\ref{eq:U3})], or other particular gates. Keeping a count of different types of gates can be rather cumbersome, especially for large circuits and when considering circuits being transformed through the application of gate identities. Furthermore, the major source of noise is generally the two-qubit gates, which is why we chose not to use counts of particular single-qubit gates.

Features like entropy (which measures the structural uncertainty or complexity of information flow within the circuit DAG) and average centrality (which captures how strongly computation is concentrated around specific operations) were consistently one of the least important features across different models when we measured that quantity like in \figref{fig:FeatureImportance} in \secref{sec:TestingOnHardware}. For that reason, these features were not included in our feature vector.

From the DAG, one can compute features like the average path length, which characterizes the typical dependency depth between operations in the quantum circuit, and the total number of directed paths in the DAG, which quantifies the number of distinct computational routes through which quantum information may flow from circuit inputs to outputs. However, both of these features have a pretty high computational cost since one needs to sweep over all the possible connections between all possible nodes, which can quickly become intractable for large quantum circuits.

Finally, we note that some features are already implicitly taken into account in the feature vector since they are closely related to or part of the definition of features there. For example, the average degree is part of the program communication [see \eqref{eq:ProgramCommunication}].


\section{Other methods of encoding data on coherence times}
\label{app:coherence}

In \secref{sec:AddingHardwareData}, we described how we added information about the coherence times of individual qubits in the feature vector by estimating failure rates for the gates in the circuit due to decoherence. Here, we discuss other methods for encoding information about the coherence times in the feature vector.

One of the simplest methods for encoding the coherence times is to directly use them as features in the feature vector (direct encoding). While this allows the hardware data to be encoded directly, it does not provide any information about qubit locations, gates, or other indicators of quantum circuit features. In other words, the connection to the circuit and processor topologies is missing.

Another method of encoding the coherence times in a compressed form would be to encode the mean coherence times (along with their standard deviations) into the feature vectors. However, this also fails to capture whether the gates in the quantum circuit are mostly performed on high- or low-quality qubits.

\begin{figure}
    \centering
    \includegraphics[width=\linewidth]{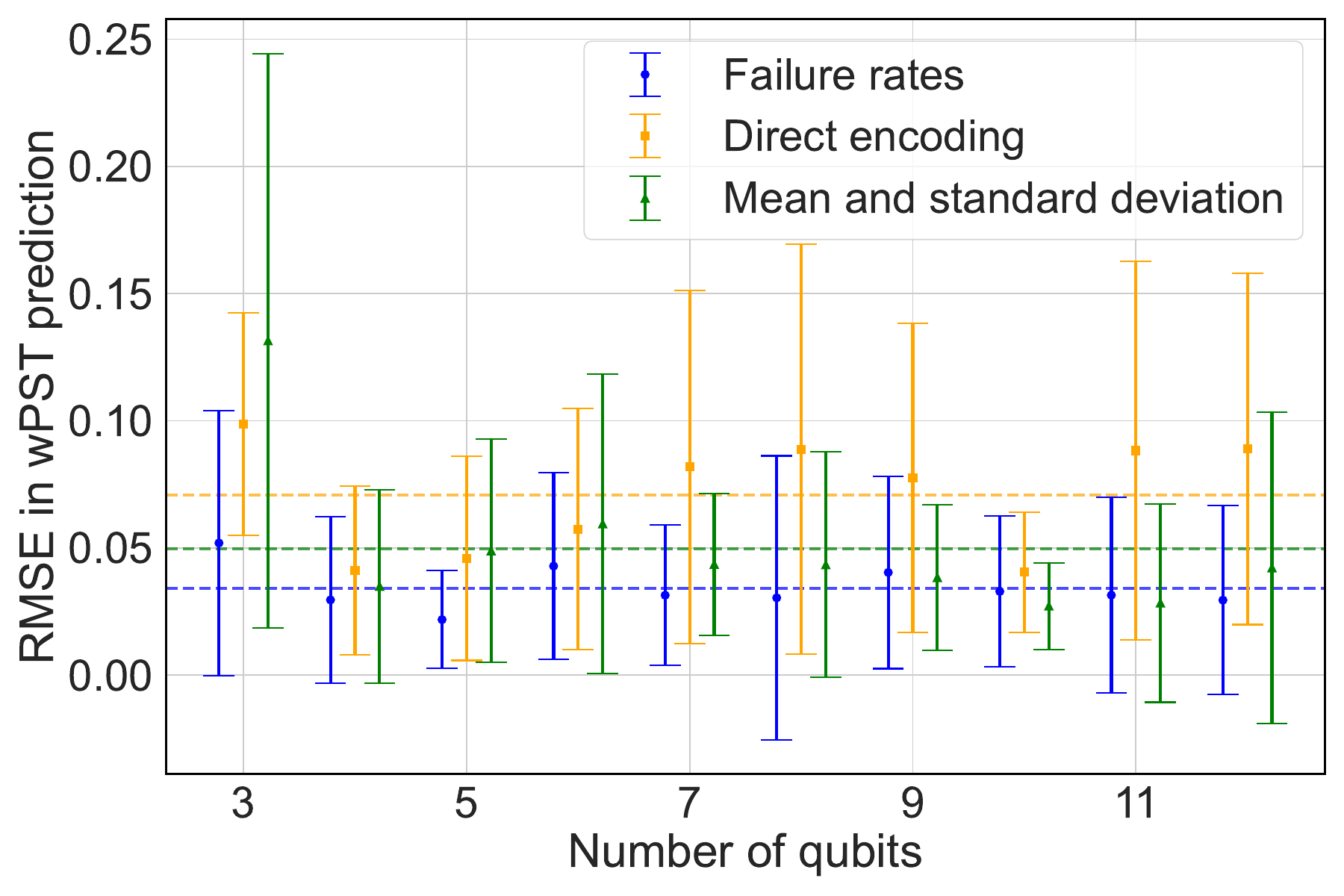}
    \caption{Performance (in terms of RMSE) in wPST prediction for three different methods for encoding information about coherence times in the feature vector, with error bars showing $\pm 1$ standard deviation. The dashed lines show the mean RMSE of each method across different qubit numbers.}
    \label{fig:rmsecoherence}
\end{figure}

In \figref{fig:rmsecoherence}, we compare the performance of the two methods discussed here with the one used in the main text, referred to as ``failure rates''. The figure shows the RMSE in wPST prediction with the three methods for different numbers of qubits. It is clear from the figure that the failure rates method performs better than the other two.


\section{Other methods of encoding coupling maps}
\label{app:coupling}

In \secref{sec:2StepDetails}, we showed how we chose to encode information about the coupling map of the quantum hardware in the two-step process for predicting the wPST of non-transpiled circuits. There, we used the coupling-degree vector for the encoding. Here, we describe two other methods for such encoding, and compare their performance with that of the method we used in the main text.


\subsection{Adjacency matrix}

Let $G = (V, E)$ be a graph representing the connectivity of a quantum device, where $V$ is the set of $n$ qubits ($|V| = n$) and $E$ is the set of allowed two-qubit interactions. The adjacency matrix $\mathbf{A} \in \mathbb{R}^{n \times n}$ then encodes the connectivity between all qubits in a compact form.
It is defined as
\begin{equation}
A_{ij} = 
\begin{cases}
    1, & \text{if qubits $i$ and $j$ are connected} \\
    0, & \text{otherwise.}
\end{cases}
\end{equation}
Here $i,j \in \{0, 1, \dots, n-1\}$. In this implementation, the adjacency matrix is made \textit{symmetric} to treat the graph as undirected:
\begin{equation}
A_{ij} = A_{ji} = 1 \quad \forall (i,j) \in E .
\end{equation}
%


\subsection{Graph features}

The following features from the coupling graph $G$ were used as a feature vector:
\begin{enumerate}
    \item Number of edges:
    \begin{equation}
        f_1 = |E|
    \end{equation}
    This is the total number of connections in the coupling map, reflecting the connectivity of the hardware.

    \item Mean degree:
    \begin{equation}
        f_2 = \frac{1}{n} \sum_{v \in V} \deg(v) ,
    \end{equation}
    where $\deg(v)$ is the total degree (sum of in-degree and out-degree) of node $v$. This measures the average number of connections per qubit.

    \item Maximum degree: 
    \begin{equation}
        f_3 = \max_{v \in V} \deg(v) .
    \end{equation}
    This comes from the most connected qubit in the device, capturing potential bottlenecks or hubs in the connectivity.

    \item Number of strongly connected components ($f_4$): \\ 
    A strongly connected component is a maximal subset of qubits $C \subseteq V$ such that for every pair $u, v \in C$, there is a directed path from $u$ to $v$. This feature captures the overall directed connectivity of the device.

    \item Graph diameter:
    \begin{equation}
        f_5 =
        \begin{cases}
            \text{diameter}(UG), & \text{if } UG \text{ is connected} \\
            -1, & \text{otherwise.}
        \end{cases}
    \end{equation}
    Here, $UG$ is the undirected version of $G$, and the diameter is defined as
    \begin{equation}
        \text{diameter}(UG) = \max_{u,v \in V} d(u,v) ,
    \end{equation}
    where $d(u,v)$ is the shortest path length between the nodes $u$ and $v$. If the graph is disconnected, the diameter is set to $-1$.

    \item Average shortest path length:
    \begin{equation}
        f_6 =
        \begin{cases}
            \frac{1}{|V|(|V|-1)} \sum_{u \neq v} d(u,v), & \text{if } UG \text{ is connected} \\
            -1, & \text{otherwise.}
        \end{cases}
    \end{equation}
This feature captures the typical distance between qubits in the hardware and reflects how efficiently two-qubit gates can be implemented across the device.
\end{enumerate}
%


\subsection{Comparison of performance}

\begin{figure*}
    \centering
        \centering
        \includegraphics[width=\textwidth]{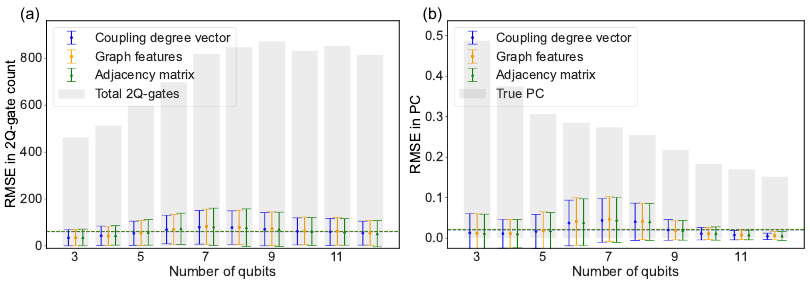} 
        \caption{Performance of the three coupling-map-encoding methods (different colors). (a) Root mean square error in prediction of the number of two-qubit gates, for different numbers of qubits on a 12-qubit simulator. The gray bars show the average total two-qubit gate count and the dashed lines show the average RMSE across different qubit numbers. The error bars show $\pm 1$ standard deviation. (b) Same as in (a), but for program communication (PC) instead of the number of two-qubit gates.}
        \label{fig:rmseApp}

\end{figure*}

In \figpanel{fig:rmseApp}{(a)} and~\figpanel{fig:rmseApp}{(b)}, we plot the RMSE for the three different coupling-map-encoding methods in the prediction of two-qubit gate counts and program communication. These two features are the most important to get right for the wPST prediction in the next step to be accurate (cf.~\figref{fig:FeatureImportance}). We see that all three methods perform similarly, with the coupling-degree-vector method perhaps being just marginally better.

However, the three methods differ in how easy their input is to compute. In the case of the adjacency matrix method, the number of features to be encoded scales as $n \times n$ (the size of the matrix for $n$ qubits). For the graph-features method, all the features need to be extracted from the coupling map. In contrast, finding the coupling-degree vector is straightforward, requires no excessive computation, and scales linearly with the number of qubits.


\section{Results for other machine learning models}
\label{app:OtherModels}

In \secref{sec:MLModel}, we described how we tested several machine learning models for prediction of PST and wPST before settling on XGBoost as the one we used in the main text. Here, in Table~\ref{tab:model_comparison}, we present the results of that testing, comparing several tree-based ensemble regression models for wPST prediction. The evaluated metrics include training time, prediction time, coefficient of determination ($R^2$), root mean squared error (RMSE), and the Pearson correlation coefficient (PCC). 

Although LightGBM slightly outperforms XGBoost in terms of $R^2$ and RMSE, the differences are minor. We ultimately selected XGBoost as the final model due to its superior balance between prediction accuracy, computational efficiency, and robustness. Specifically, XGBoost achieves the fastest training and prediction times while maintaining competitive accuracy and the highest correlation with ground-truth values. This trade-off makes XGBoost particularly suitable for scalable applications and repeated model evaluation scenarios, which will be critical when this model is used within a quantum compiler.

\begin{table*}
\centering
\caption{Performance comparison of tree-based regression models. The best performer for each category is highlighted in bold.}
\label{tab:model_comparison}
\renewcommand{\arraystretch}{1.1}   
\setlength{\tabcolsep}{4pt}          
\begin{tabular}{l|ccccc}
\toprule
Model & Training & Prediction & $R^2$ & RMSE & PCC \\
 & time (s) & time (s) &  &  &  \\
\midrule
RandomForest~\cite{Breiman2001}   & 5.670 & 0.040 & 0.8998 & 0.1085 & 0.9487 \\
ExtraTrees~\cite{Geurts2006}     & 3.539 & 0.046 & 0.8873 & 0.1150 & 0.9423 \\
GBDT~\cite{friedman2001greedy}          & 1.798 & 0.004 & 0.9061 & 0.1050 & 0.9520 \\
Bagging~\cite{Breiman1996}        & 5.749 & 0.052 & 0.8999 & 0.1084 & 0.9487 \\
AdaBoost~\cite{FREUND1997119}       & 0.433 & 0.003 & 0.8851 & 0.1161 & 0.9460 \\
XGBoost~\cite{10.1145/2939672.2939785}        & \textbf{0.072} & \textbf{0.001} & 0.9068 & 0.1046 & \textbf{0.9530} \\
LightGBM~\cite{NIPS2017_6449f44a}       & 0.094 & 0.003 & \textbf{0.9080} & \textbf{0.1039} & 0.9524 \\
CatBoost~\cite{dorogush2018catboostgradientboostingcategorical}       & 0.180 & 0.012 & 0.9029 & 0.1068 & 0.9504 \\
RandomJungle~\cite{Schwarz2010}   & 1.609 & 0.109 & 0.8926 & 0.1124 & 0.9449 \\
\bottomrule
\end{tabular}
\end{table*}


\section{wPST and PST scaling under independent bit-flip noise}
\label{app:bitflip}

The simplest noise model that exposes the structural difference and scaling behavior of wPST and PST for large system sizes is that of independent bit flipping. Under independent bit-flip noise, each of the $n$ measured bits is flipped with probability $\epsilon$. The measured bits $x_q$ are thus independent and identically distributed (i.i.d.)\ Bernoulli variables with $\Pr[x_q = 1] = \epsilon$. This model is deliberately crude and blends gate errors, decoherence, and readout error into a single effective per-qubit rate; it also ignores specific circuit features that can turn bit-flip errors into phase-flip errors. We use this model here to derive exact expressions, show justification that the thresholded and unthresholded wPST behave similarly, and provide an analysis of the shot budget for wPST and PST.


\subsection{Exact values}

Since PST is the probability of the single all-zero outcome, we can factorize the joint distribution for independent noise:
\begin{equation}
    \mathrm{PST} = \Pr\left[ \textstyle\bigwedge_{q=1}^{n} x_q = 0 \right]
    = \prod_{q=1}^{n} \Pr[x_q = 0]
    = (1-\epsilon)^n .
    \label{eq:pstbitflipexact}
\end{equation}
For wPST we use the marginal representation in \eqref{eq:wpstmarginals} and set the single-qubit marginal as $m_q = \Pr[x_q = 0] = 1-\epsilon$ independently of $q$, yielding
\begin{equation}
    \mathrm{wPST}_{w_{\rm th}=0} = \frac{1}{n}\sum_{q=1}^{n} m_q = 1-\epsilon .
    \label{eq:wpstbitflfipexact}
\end{equation}
The two expressions differ in exactly one aspect: Equation~(\ref{eq:pstbitflipexact}) is the probability that all $n$ random variables succeed \emph{simultaneously}, and therefore multiplies. However, \eqref{eq:wpstbitflfipexact} is the \emph{expectation of their mean}, and therefore carries no $n$-dependence whatsoever. This is the analytical content of the linear-versus-multiplicative statement made in \secref{sec:wpst}, and it predicts the curves observed in \figref{fig:wpstpst}.


In the Hamming-weight representation in \eqref{eq:wpsthamming}, we can derive the same result as a consistency check, as the number of zeros $k$ is binomially distributed,  $k \sim \mathrm{Bin}(n, 1-\epsilon)$, so that the shell probabilities in \eqref{eq:shell} are
\begin{equation}
    Q(k) = \binom{n}{k} (1-\epsilon)^{k} \epsilon^{\,n-k} ,
\end{equation}
and consequently
\begin{equation}
    \mathrm{wPST}_{w_{\rm th}=0}
    = \frac{1}{n}\sum_{k=0}^{n} k\, Q(k)
    = \frac{1}{n}\,\mathbb{E}[k]
    = 1-\epsilon ,
\label{eq:wPSTwTH0}
\end{equation}
in agreement with \eqref{eq:wpstbitflfipexact}.  Setting $k_{\rm th}=n$ keeps only the $k=n$ term and returns $Q(n)=(1-\epsilon)^n = \mathrm{PST}$, as required.



\subsection{Effect of the threshold}

In practice, we use $w_{\rm th}=0.5$ and remove the shells with $k < k_{\rm th} = \lceil n/2 \rceil$ from the sum in \eqref{eq:wPSTwTH0} above:
\begin{equation}
    \mathrm{wPST}_{w_{\rm th}=0.5}
    = (1-\epsilon) - \frac{1}{n}\sum_{k=0}^{k_{\rm th}-1} k\, Q(k) .
    \label{eq:threshcorrection}
\end{equation}
The correction term is bounded by the probability of the discarded shells, $\Pr[k<n/2]$, which for $\epsilon < 1/2$ is exponentially small in $n$ by the Chernoff bound:
\begin{equation}
    \Pr\!\left[ k < \tfrac{n}{2} \right] \le \exp\!\left[ -n\, D\!\left( \tfrac{1}{2} \,\middle\|\, 1-\epsilon \right) \right] ,
\end{equation}
with $D(a\|b)$ the binary Kullback--Leibler divergence. Provided $\epsilon$ is sufficiently smaller than $1/2$, so that $D(\tfrac12\|1-\epsilon)$ stays finite, the discarded shells carry exponentially little probability and \eqref{eq:wpstbitflfipexact} holds for $\mathrm{wPST}_{w_{\rm th}=0.5}$ up to corrections exponentially small in $n$. The threshold thus plays no role in the noise scaling, which follows from the additivity of \eqref{eq:wpstmarginals} alone. The threshold's purpose is interpretational, refusing partial credit to majority-incorrect outcomes.



\subsection{Shot budget}

This section discusses an operational reason to prefer wPST over the standard PST, which is the precision with the shot budget. 
Let $N$ be the number of shots for a quantum circuit execution. 

\subsubsection{PST}

The PST estimator can be written as
\begin{equation}
 \widehat{\mathrm{PST}} = n_{0^n}/N
\end{equation}
and follows a binomial distribution with success probability $p = \rm PST$.
The variance of the estimator can be written as
\begin{equation}
\mathrm{Var}\!\mleft[\widehat{\mathrm{PST}}\mright]
=
\frac{p(1-p)}{N} .
\end{equation}

The relative standard error (RSE) $\delta$ is defined as the standard deviation divided by the mean,
\begin{equation}
    \delta = \frac{\sigma[\widehat{\mathrm{PST}}]}{\mathbb{E}[\widehat{\mathrm{PST}}]}.
\end{equation}
Since
\begin{equation}
    \mathbb{E}[\widehat{\mathrm{PST}}] = p,
\end{equation}
we have
\begin{equation}
     \delta
     = \sqrt{\frac{1-p}{N\,p}}.
 \end{equation}
For $p \ll 1$, the $(1-p)\approx 1$ factor drops out, leaving
\begin{equation}
     \delta
     \approx \frac{1}{\sqrt{N\,p}}.
 \end{equation}
Achieving a fixed relative precision $\delta$ therefore requires
 \begin{equation}
     N \;\sim\; \frac{1}{\delta^{2}\,p} ,
 \end{equation}
and since $p = (1-\epsilon)^n$, the number of shots required scales as
 \begin{equation}
     N \sim \mathcal{O} \mleft( \frac{1}{\delta^{2}(1-\epsilon)^n} \mright) ,
 \end{equation}
 which is exponential in the number of qubits. Equivalently, at a fixed shot budget $N$, the relative error itself grows exponentially, $\delta \sim (1-\epsilon)^{-n/2}/\sqrt{N}$: no realistic budget keeps pace with system size.
 
 At the same time, the absolute error itself is not dependent on $n$:
 \begin{equation}
     \sigma[\widehat{\mathrm{PST}}] \le 1/(2\sqrt{N})
 \end{equation}
 for any $N$, but that is not sufficient here. Since PST becomes exponentially small with system size, an absolute error tolerance does not characterize the usefulness of the estimate. Instead, one must resolve differences relative to the value itself, making the relative standard error the relevant quantity. Therefore, it fails to separate a well-compiled circuit from a badly compiled one once both have $\mathrm{PST}$ below that range. 

\subsubsection{wPST}

Now we consider the same problem with wPST. For wPST, the result for a single shot $s$ can be written as
 \begin{equation}
     W_s = \frac{1}{n}\sum_{q=1}^{n} \mathbbm{1}[x_{s,q}=0] \in [0,1]
 \end{equation}
and one can write the wPST ($w_{th} = 0$) as
\begin{equation}
  \widehat{\mathrm{wPST}}_0 = \frac{1}{N}\sum_{s=1}^{N} W_s ,
  \qquad \mathbb{E}[W_s] = 1-\epsilon . 
\end{equation}

Since $W_s$ is bounded in $[0,1]$, Popoviciu's inequality~\cite{popo} gives $\mathrm{Var}[W_s] \le 1/4$, and hence
\begin{equation}
 \sigma[\widehat{\mathrm{wPST}}_0] \le \frac{1}{2\sqrt{N}} ,  
\end{equation}
which holds for any noise model, correlated or not.
For the independent error model chosen here, the exact results can be computed:
 \begin{align}
     \mathrm{Var}[W_s] &= \frac{1}{n^{2}} \sum_{q=1}^{n} \mathrm{Var}\mleft[\mathbbm{1}[x_q=0]\mright]
     = \frac{\epsilon(1-\epsilon)}{n} , \\
     \sigma[\widehat{\mathrm{wPST}}_0] &= \sqrt{\frac{\epsilon(1-\epsilon)}{N n}} = \mathcal{O} \mleft( \frac{1}{\sqrt{N n}} \mright) .
 \end{align}
In this case, we then obtain
\begin{equation}
\delta = 
    \frac{\sigma[\widehat{\mathrm{wPST}}_0]}
    {\mathbb{E}[\widehat{\mathrm{wPST}}_0]}
    =
    \sqrt{\frac{\epsilon}{Nn(1-\epsilon)}}
    =
    \mathcal{O} \mleft(\frac{1}{\sqrt{Nn}}\mright).
\end{equation}
Achieving a fixed relative precision $\delta$ therefore requires
\begin{equation}
    N
    \sim
    \frac{\epsilon}{\delta^2 n(1-\epsilon)}
    =
    \mathcal{O} \mleft(\frac{1}{n\delta^2}\mright).
\end{equation}

Model-independently, fixed absolute precision on wPST thus costs $N=\mathcal{O}(1)$ shots, regardless of $n$ and of the noise model. Under the independent-error model, the cost even decreases to $\mathcal{O}(1/n)$: each shot contributes $n$ informative bits rather than the single bit PST extracts from $n_{0^n}$.
The asymmetry of exponential shots requirement for PST against constant shots for wPST is an operational advantage as an FoM, especially when building compilers for larger quantum systems. This makes wPST substantially more practical as a figure of merit for benchmarking and optimizing large-scale quantum compilers.




\bibliography{biblio.bib}

\end{document}